\documentclass[longauth]{aa}  
\usepackage{graphicx}
\usepackage[dvipsnames]{xcolor}
\usepackage[normalem]{ulem}
\usepackage{amsmath,amssymb}
\usepackage{txfonts}
\DeclareUnicodeCharacter{0308}{\'{e}}

\begin{document}

   \title{FAUST IX. Multiband, multiscale dust study of L1527 IRS}

   \subtitle{Evidence for variations in dust properties within the envelope of a class 0/I young stellar object}

  \author{L. Cacciapuoti
          \inst{1,2,3}
          \and E. Macias
          \inst{1}
          \and A. J. Maury
          \inst{4}
          \and C. J. Chandler
          \inst{5}
          \and N. Sakai
          \inst{6}
          \and $\L$. Tychoniec
          \inst{1}
          \and S. Viti
          \inst{7,8}
          \and A. Natta
          \inst{9}
          \and{M. De Simone}
          \inst{1,2}
          \and A. Miotello
          \inst{1}
          \and C. Codella
          \inst{2,10}
          \and C. Ceccarelli 
          \inst{10}
          \and L. Podio
          \inst{2}
          \and D. Fedele
          \inst{2}
          \and D. Johnstone
          \inst{11,12}
          \and Y. Shirley 
          \inst{13}
          \and B. J. Liu
          \inst{14,15}
          \and E. Bianchi
          \inst{16}
          \and Z. E. Zhang
          \inst{5}
          \and J. Pineda
          \inst{17}
          \and L. Loinard
          \inst{18}
          \and F. Ménard
          \inst{9}
          \and U. Lebreuilly
          \inst{4}
          \and R. S. Klessen
          \inst{19,20}
          \and P. Hennebelle 
          \inst{4}
          \and S. Molinari
          \inst{21}       
          \and L. Testi
          \inst{2,22}
          \and S. Yamamoto
          \inst{23}
          }

\titlerunning{FAUST X. Multiscale dust continuum of L1527}
\authorrunning{Cacciapuoti et al.}

  \institute{
         European Southern Observatory, Karl-Schwarzschild-Strasse 2 D-85748 Garching bei Munchen, Germany
         \and 
         INAF, Osservatorio Astrofisico di Arcetri, Largo E. Fermi 5, I-50125, Firenze, Italy
         \and
        Fakultat fur Physik, Ludwig-Maximilians-Universitat Munchen, Scheinerstr. 1, 81679 Munchen, Germany
         \and
         Universit\'{e} Paris-Saclay, Universit\'{e} Paris Cité, CEA, CNRS, AIM, 91191, Gif-sur-Yvette, France
         \and
         National Radio Astronomy Observatory, PO Box O, Socorro, NM 87801, USA
         \and
         RIKEN Cluster for Pioneering Research, 2-1, Hirosawa, Wako-shi, Saitama 351-0198, Japan
         \and
        Leiden Observatory, Leiden University, PO Box 9513, 2300 RA Leiden, The Netherlands
         \and
         Department of Physics and Astronomy, University College London, Gower Street, London, WC1E 6BT, UK
         \and
        Dublin Institute for Advanced Studies (DIAS), School of Cosmic Physics, Astronomy and Astrophysics Section, 31 Fitzwilliam Place, Dublin 2, Ireland
         \and
         Univ. Grenoble Alpes, CNRS, IPAG, 38000 Grenoble, France
        \and
        NRC Herzberg Astronomy and Astrophysics, 5071 West Saanich Rd, Victoria, BC, V9E 2E7, Canada
         \and
         Department of Physics and Astronomy, University of Victoria, Victoria, BC, V8P 5C2, Canada
         \and 
        Steward Observatory, 933 North Cherry Avenue, Tucson, AZ 85721, USA
         \and
        Physics Department, National Sun Yat-Sen University, No. 70, Lien-Hai Road, Kaosiung City 80424, Taiwan, R.O.C.
         \and 
         Institute of Astronomy and Astrophysics, Academia Sinica, 11F of Astronomy-Mathematics Building, AS/NTU No.1, Sec.4, Roosevelt Rd, Taipei 10617, Taiwan, ROC.
         \and
         Excellence Cluster ORIGINS, Boltzmannstraße 2, D-85748 Garching bei Mu ̈nchen, Germany
        \and 
         Max-Planck-Institut für extraterrestrische Physik (MPE), Giessenbachstr. 1, D-85741 Garching, Germany
         \and
         Instituto de Radioastronomía y Astrofísica, Universidad Nacional Autónoma de México Apartado 58090, Morelia, Michoacán, Mexico
         \and
         Universität Heidelberg, Zentrum für Astronomie, Institut für Theoretische Astrophysik, Albert-Ueberle-Straße 2, 69120 Heidelberg, Germany
         \and
         Universität Heidelberg, Interdisziplinäres Zentrum für Wissenschaftliches Rechnen, Im Neuenheimer Feld 205, 69120 Heidelberg, Germany
        \and
        INAF-Istituto di Astrofisica e Planetologia Spaziali, Via del Fosso del Cavaliere 100, I-00133, Rome, Italy
        \and
        Dipartimento di Fisica e Astronomia "Augusto Righi" Viale Berti Pichat 6/2, Bologna
        \and 
        Department of Physics, The University of Tokyo, 7-3-1, Hongo, Bunkyo-ku, Tokyo 113-0033, Japan
             }

   \date{Received 21 February, 2023; accepted 5 June, 2023}

\abstract{
  {\textit{Context.}}
   Early dust grain growth in protostellar envelopes infalling on young disks has been suggested in recent studies, supporting the hypothesis that dust particles start to agglomerate already during the class 0/I phase of young stellar objects (YSOs). If this early evolution were confirmed, it would impact the usually assumed initial conditions of planet formation, where only particles with sizes $\lesssim 0.25 \mu$m are usually considered for protostellar envelopes. 
   
   {\textit{Aim.}} 
   We aim to determine the maximum grain size of the dust population in the envelope of the class 0/I protostar L1527 IRS, located in the Taurus star-forming region (140 pc).

   {\textit{Methods.}} 
   We use Atacama Large millimeter/submillimeter Array (ALMA) and Atacama Compact Array (ACA) archival data and present new observations, in an effort to both enhance the signal-to-noise ratio of the faint extended continuum emission and properly account for the compact emission from the inner disk. Using observations performed in four wavelength bands and extending the spatial range of previous studies, we aim to place tight constraints on the spectral ($\alpha$) and dust emissivity ($\beta$) indices in the envelope of L1527 IRS.

   {\textit{Results.}} 
   We find a rather flat $\alpha \sim$ 3.0 profile in the range 50-2000 au. 
   Accounting for the envelope temperature profile, we derived values for the dust emissivity index, $0.9 < \beta < 1.6$, and reveal a tentative, positive outward gradient. This could be interpreted as a distribution of mainly interstellar medium (ISM) like grains at 2000 au, gradually progressing to (sub)millimeter-sized dust grains in the inner envelope, where at R $=$ 300 au, $\beta = 1.1 \pm 0.1$. 
   
   Our study supports a variation of the dust properties in the envelope of L1527 IRS. We discuss how this can be the result of in situ grain growth, dust differential collapse from the parent core, or upward transport of disk large grains.
}
             
\maketitle
%

\section{Introduction}
\label{sec:intro}
Circumstellar disks orbiting class II young stellar objects (YSOs) are also commonly referred to as protoplanetary disks even though a large fraction of them already display structures thought to be shaped by embedded planets \citep{Andrews2018}. Rings, gaps, and spirals have also been observed in younger (<1Myr) disks (\citealt{AlmaPartenership2015}, \citealt{Sheehan2018}, \citealt{Seguracox2020}, \citealt{Nakatani2020}).
Several mechanisms have been proposed to explain the formation of structures during the early stages of the disk lifetime, including gravitational instabilities \citep{Takahashi2014}, disk winds (\citealt{Johansen2009}, \citealt{Takahashi2018}), and the evolution of dust \citep{Okuzumi2016}. Along with these possibilities, the early formation of large planetesimals that gravitationally interact with the disk remains a viable explanation for the observations. 

The early formation of planetary embryos is also suggested by the dust mass budget of evolved disks when compared to their younger progenitors (\citealt{Testi2014}, \citealt{Ansdell2016}). \citet{Manara2018} estimated that the solid (dust) mass observed in evolved class II disks of the Ophiuchus \citep{Sanchis2020} and Lupus star-forming-region is 1 or 2 orders of magnitude lower than the mass of the expected exoplanet population. Conversely, the class 0/Is disks of Perseus contain roughly the same mass in solids as the known exoplanets (\citealt{Williams2019}, \citealt{Tychoniec2020}. One possible explanation is that the missing mass in the class II disks is the early conversion of class 0/I solids into planetesimals (e.g. \citealt{Testi2022}, \citealt{Bernabo2022}, \citealt{Xu2023}). 
In a separate study, \citet{Mulders2021} suggest that the solid masses in class II disks and in exoplanets might be of the same order of magnitude. Nevertheless, an unrealistic planet formation efficiency of 100\% would be required to place the beginning of planet formation during the class II stage. 

Direct observational evidence for early-formed, kilometer-sized planetesimals is out of reach for any radio interferometer and thus this problem remains open. Furthermore, observational constraints on the initial properties of the dust population are needed for simulations to reconstruct the pathways that lead to planetesimal formation, starting from submicron-sized interstellar medium (ISM) dust particle interactions in a core accretion paradigm (e.g., \citealt{Safronov1969}, \citealt{Goldreich1973}, 
\citealt{Ormel2009}, \citealt{Birnstiel2016}, \citealt{Draz2022}). Even more importantly in this context, in an effort to investigate planetesimal formation at early stages, \citet{Cridland2022} find that differential gas and dust replenishment of a protoplanetary disk from its surrounding envelope sets favorable conditions for planetesimal formation. However, grains that are moderately coupled to the gas ($>$30 $\mu$m) are required to meet the streaming instability conditions that lead to the birth of these planetary embryos \citep{Youdin2005}.
Moreover, constraining grain sizes during the protostellar collapse phase has proven to be important for understanding the role of magnetic breaking during the early stages of disk formation. Since small grains are the main charge carriers, their distribution is paramount for the coupling of the infalling material with magnetic fields. In turn, such a coupling seems to regulate disk masses and sizes (\citealt{Zhao2016}, \citealt{Lebreuilly2020}).

An additional reason to investigate the properties of dust grains in these extended envelopes is to constrain the role that dust grains play as sites - and catalysts - of molecular reactions in diffuse environments. The most commonly accepted pathways to produce complex organic molecules all rely on the presence of icy mantles on dust grains to capture the building blocks of these complex molecules and facilitate reactions among them (e.g., \citealt{Tielens1982}). 
\begin{table}
\caption{Main properties of our target, L1527 IRS. The age of the object has been roughly estimated by [1] \citep{Tobin2012} based on the mass loss rate of the source.
The mass of the protostar has been estimated by means of kinematical analysis by [1] and [2] \citep{Aso2017}. The bolometric luminosity was reported by [3] \citep{Karska2018} and might suffer up to a factor two uncertainty due to the high inclination of the disk (e.g., \citealt{Whitney2003}).The envelope mass was derived in [4] \citep{Motte2001} based on the IRAM 30 m telescope and MPIfR bolometer arrays' 1.3 millimeter maps.}
\label{tab:l1527}
\centering
\begin{tabular}{lccc}
\hline \\
            Alternative ID  &  IRAS 04368+2557 \\ \\
            RA & 04h 39m 53.88s \\ \\
            Dec & +26$^{\circ}$ 03'09$\farcs$56 \\ \\
            Age [yr] & < 3 $\cdot$ $10^5$$^{[1]}$ \\ \\
            $M_{*} [M_{\odot}]$ & 0.2$^{[1]}$-0.45$^{[1,2]}$ \\ \\
            $L_{bol} [L_{\odot}]$ & 1.6$^{[3]}$  \\ \\
            $M_{4200au}^{env} [M_{\odot}]$ & 0.8$^{[4]}$  \\ \\
\hline
\end{tabular}
\end{table}

One way to probe dust grain properties in protostellar environments relies on estimates of their spectral index ($\alpha$), the slope of the spectral energy distribution (SED) across (sub)millimeter wavelengths. 
Specifically, if dust opacity scales as $\kappa \propto \nu^{\beta}$ and if the Rayleigh-Jeans (RJ) approximation holds, $\beta = \alpha - 2$ in the optically thin regime (e.g., \citealt{BeckwithSargent1991}, \citealt{MiyakeNakagawa1993}, \citealt{Natta2007}). 
Typical spectral indices for the ISM - $\beta \sim 1.7$ - correspond to grain sizes in the range 
100 \AA\ -- 0.3 $\mu$m \citep{Weingartner2001}. On the other hand, $\beta < 1$ has been observed in class II objects, suggesting the presence of larger grains ($a \geq 1$ mm) in more evolved disks (\citealt{BeckwithSargent1991} ,\citealt{Testi2003}, \citealt{Ricci2010}, \citealt{Testi2014}, \citealt{Tazzari2021}).

In the recent past, several works have attempted to measure the dust emissivity index $\beta$ at the large scales of class 0/I protostellar envelopes to constrain whether dust growth might be significant at these very early stages of star and planet formation. Many have found surprisingly low $\beta$ values ($\leq 1$) and interpreted this result as evidence for early dust growth (\citealt{Kwon09}, \citealt{Shirley2011}, \citealt{Chiang2012}, \citealt{Miotello2014}, \citealt{LG2019}, \citealt{Valdivia2019}). Following similar methods, other works have not found hints of such growth \citep{AG2019}. And, in their sample study of ten CALYPSO \citep{Maury2019} class 0/I sources, \citet{Galametz2019} found examples of both relatively low and large $\beta$ values. 

The hypothesis of early grain growth in the envelopes of class 0/Is has been considered to be challenging from a theoretical perspective: growth to millimeter-sized particles seems to require environments characterized by higher density and/or longer timescales than the average class 0/I envelope (e.g., \citealt{Ormel2009}, \citealt{Guillet2020}, \citealt{Lebreuilly2023}, \citealt{Bate2022} and \citealt{Silsbee2022}). Using both analytical models and numerical simulations, these authors find that dust particles cannot grow larger than $\sim 2 \mu$m in collapsing envelopes. It will be crucial for next simulations to incorporate generally disregarded effects, like the dust back-reaction on the turbulence through gas-dust friction and dust-magnetic-field interaction.
Early growth is not the only possible explanation for the mentioned observations. \citet{Lebreuilly2020} proposed a scenario in which the differential collapse of dust grains of different mass through the prestellar core leads to a stratification of the dust sizes since larger grains collapse faster.
Finally, another possibility could be an uplifting of grown disk grains to the inner envelope by the protostellar outflows and/or jets (\citealt{Wong2016}, \citealt{Tsukamoto2021}).

It is thus imperative to characterize dust properties and evolution in the very early stage of star and planet formation.

\begin{table*}[t]
\footnotesize
\centering   
\caption{ALMA observations. For projects with more than one execution block, we only report the total integration time of the project. The rescaling factors are relative to the dataset whose factor is exactly 1.000.}             
\label{table:1}      
\begin{tabular}{c c c c c c c c}     
\hline\hline       
Project Code & P.I. & Date &  Integration (s) & Resolution & Frequency (GHz) & Rescale Factor & CASA version  \\ 
\hline 
    & & & Band 3 & & &   \\     \hline       
    2015.1.00261.S & Ceccarelli, C. & 01/03/2016 & 3113  & 2$\farcs$ & 85-87 & 0.93 & 4.5.2\\ 
                   &                & 02/03/2016 &  & 2$\farcs$4 & 90-92 & 0.92 & 4.5.2 \\
    2016.1.01245.S & Cox, E.        & 04/01/2017 & 907   & 2$\farcs$1  & 99-114 & 1.06 & 4.7.0-1 \\
    2016.1.01541.S & Harsono, D.    & 21/12/2016 & 635   & 1$\farcs$8  & 92-105 & 1.14 & 4.7.0 \\
    2017.1.00509.S & Sakai, N.      & 14/11/2017 & 6150  & 0$\farcs$09 & 85-99 & 1.000 & 5.1.1-5\\
                   &                & 14/11/2017 &       & 0$\farcs$09 & 85-99 & 1.04 & 5.1.1-5\\
                   &                & 14/11/2017 &       & 0$\farcs$09 & 85-99 & 1.04 & 5.1.1-5\\
    2018.1.01205.L & Yamamoto, S.    & 4/01/2017  & 1783  & 1$\farcs$1  & 93-108& - & 5.6.1-8 \\
\hline 
& & & Band 4 & & &  \\ \hline    
    2016.1.01203.S & Oya, Y.     & 19/11/2016 & 4697 & 0$\farcs$8 & 138-150 & 0.96 & 5.1.1 \\
                   &             & 03/09/2017 &      & 0$\farcs$1 & 138-150 & 0.92 & 5.1.1 \\
    2016.1.01541.S & Harsono, D. & 10/03/2017 & 241  & 2$\farcs$1 & 144-154 & 0.97 & 4.7.0\\
                   &             & 31/03/2017 &      & 2$\farcs$1 & 144-154 & 1.000 & 4.7.0\\
\hline 
& & & Band 6 & & &  \\ \hline                  
    2012.1.00647.S & Ohashi, N. & 20/07/2014 & 2117 & 0$\farcs$4     & 218-233 & 1.000 & 4.2.1\\
                   &            & 20/07/2014 &      & 0$\farcs$4     & 218-233 & 1.01 & 4.2.1\\
    2013.1.01086.S & Koyamatsu, S. & 24/05.2015 & 2902 & 0$\farcs$6  & 219-234 & 0.85 & 4.5.0\\
                   &               & 20/09/2015  &      & 0$\farcs$2 & 219-234 & 0.99 & 4.5.0 \\
    2013.1.00858.S & Sakai, N. & 18/07/2015 & 2388 & 0$\farcs$2      & 245-263 & 0.96 & 4.4.0\\
    2012.1.00193.S & Tobin, J. & 11/08/2015 & 7348 & 0$\farcs$2      & 244-260 & 0.94 & 4.3.1\\
                   &           & 02/09/2015 &      & 0$\farcs$2      & 244-260 & 1.05 & 4.3.1\\
    2017.1.01413.S & van 't Hoff, M. & 10/09/2018 & 968 & 0$\farcs$3 & 226-240 & 1.05 & 5.4.0\\
                   &                 & 28/09/2018 &     & 0$\farcs$3 & 226-240 & 1.07& 5.4.0\\
    2011.0.00604.S & Sakai, N. & 10/08/2012 &  4838 & 0$\farcs$6     & 245-262 & 1.14 & 4.2.1 \\
                   &           & 26/08/2014 &       & 0$\farcs$6     & 245-262 & 1.09 & 4.2.1 \\
    2011.0.00210.S & Ohashi, N. & 26/08/2012 & 1663 & 0$\farcs$7     & 219-231 & 1.05 & 4.2.1 \\  
    2013.1.01331.S & Sakai, N. & 02/02/2015 & 1209.6  & 1$\farcs$2   & 216-235 & 0.97 & 4.3.0\\  
    2018.1.01205.L &  Yamamoto, S. & 26/10/2018 & 1360.8 & 0$\farcs$3 & 244-262 & 1.10 & 5.6.1-8  \\
                   &              & 17/03/2019 & 362.0 & 1$\farcs$1  & 244-262 & 1.10 & 5.6.1-8  \\
                   &              & 15/12/2019 &       &    1$\farcs$1  & 244-262 & 1.11 & 5.6.1-8 \\
                   &              & 26/10/2018 & 2358 & 0$\farcs$3   & 244-262 & 1.01 & 5.6.1-8 \\
                   &              & 07/03/2019 & 635.0 & 1$\farcs$1  & 244-262 & 1.004 &5.6.1-8  \\

\hline
& & & Band 7 & & &  \\ \hline                  
    2012.1.00346.S & Evans, N. & 14/06/2014 & 635  & 0$\farcs$3 & 343-357 & 1.11 &4.2.1\\
    2015.1.01549.S & Ohashi, N. & 26/07/2016 & 847 & 0$\farcs$2& 329-341 & 1.000 & 4.5.3 \\
    2016.A.00011.S & Sakai, N. & 29/07/2017 & 3592 & 0$\farcs$075 & 339-352 & 0.95 & 4.7.2\\  
                   &           & 05/09/2017 &       & 0$\farcs$075 & 339-352 & 1.01 & 4.7.2\\ 
    2011.0.00604.S & Sakai, N. & 29/08/2012 & 4233 & 0$\farcs$5 & 338-352 & 1.12 &4.2.1 \\  \hline
\end{tabular}
\end{table*}

In this paper we focus on a single source in Taurus, L1527 IRS, (140 pc, e.g., \citealt{Torres2007}, \citealt{Zucker2019}), hereafter L1527, to benchmark a dust continuum study across multiple frequencies and physical scales. 
Due to its distance and brightness, L1527 is one of the best studied class 0/I YSOs. The presence of an edge-on circumstellar disk was first suggested by \citet{Bontemps1996} and \citet{Ohashi1997} based on observations of the orientation of the bipolar outflow.
It has been later confirmed by means of Spitzer and Gemini-North scattered light observations as an edge-on dark lane obscuring the central protostar (\citealt{Tobin2008} and \citealt{Tobin2010a}, respectively) as well as high-resolution Submillimeter Array (SMA) and Combined Array for Research in Millimeter Astronomy (CARMA) interferometric observations of its continuum emission \citep{Tobin2013}. The Atacama Large Millimeter/submillimeter Array (ALMA) C$^{18}$O kinematic detection of the disk was reported by \citet{Ohashi2014} and \citet{Aso2017}. Furthermore, this edge-on, warped disk presents asymmetries at $\sim 20$ au that are consistent with spiral structures, the physical origin of which has been proposed to be gravitational instability (\citealt{Sakai2019}, \citealt{Nakatani2020}, \citealt{Ohashi2022}, \citealt{Sheehan2022}).  
Observations of L1527 also display a bipolar jets and outflows extending to $20000$ au perpendicular to the disk plane, and carving a cavity in the collapsing envelope (\citealt{Bontemps1996}, \citealt{Ohashi1997}, \citealt{Hogerheijde1998}, \citealt{Tobin2010a}, \citealt{Oya2015}). Being this object in a late stage of accretion, its molecular jet and outflow are not very energetic \citep{Podio2021}. Table \ref{tab:l1527} summarizes the main properties of this young source.

We aim to study the dust continuum emission of L1527 using the Atacama Large millimeter/submillimeter Array (ALMA) and the Atacama Compact Array (ACA) to probe the outer ($10^3$ au) and inner ($10^2$ au) envelope, as well as the disk  (10 au). We consider observations of our target in four different ALMA Bands (3, 4, 6, 7). Section 2 presents the ALMA and ACA observations that we have used throughout the analysis and the data reduction process. In Section 3, we justify the uv-plane geometrical modeling used for the source. In Section 4, we analyze the spectral and dust emissivity indices of L1527 at different scales. We discuss our results in Section 5, and wrap up our conclusions in Section 6.

\begin{table*}[t]
\footnotesize
\caption{ACA observations. The total integration times refer to the sum of the integration times within the entire project, in the cases with more than one execution block. The rescaling factors are relative to the 12m array dataset observed with the 12-m array, whose rescaling factor is 1.000 (cfr Table\ref{table:1}).}             
\label{table:2}      
\centering          
\begin{tabular}{c c c c c c c c}     
\hline\hline       
Project Code & P.I. & Date &  Integration (s) & Resolution & Frequency (GHz) & Rescaling Factor & CASA version  \\ 
\hline 

  \hline 
    & & & Band 3 & & &   \\     \hline       
     2016.1.01541.S & Harsono, D. & 30/01/2018 & 907 & 13$\farcs$0 & 92-105    & 1.06 & 4.7.0 \\  
     2018.1.00799.S & Pineda, J.  & 05/10/2018 &5235 & 12$\farcs$0 & 92-107    & 1.06 & 5.4.0\\
                    &             & 05/10/2020 &     & 12$\farcs$0 & 92-107    & 1.07 & 5.4.0\\
                    &             & 06/10/2018 &     & 12$\farcs$0 & 92-107    & 1.05 & 5.4.0\\
                    &             & 06/10/2018 &     & 12$\farcs$0 & 92-107    & 1.08 & 5.4.0\\
                    &             & 06/10/2020 &     & 12$\farcs$0 & 92-107    & 1.06 & 5.4.0\\
     
\hline 
& & & Band 4 & & &  \\ \hline    
    2016.1.01541.S & Harsono, D.  & 27/10/2016 & 302    & 9$\farcs$1 & 144-154 & 1.07 &4.7.0 \\  
    2016.2.00171.S & Harsono, D.  & 24/08/2017 &  6168  & 7$\farcs$8 & 144-154 & 1.03 &4.7.2\\
                    &             & 03/09/2017 &        & 7$\farcs$8 & 144-154 & 1.01 &4.7.2\\
                    &             & 04/09/2017 &        & 7$\farcs$8 & 144-154 & 0.93 &4.7.2\\
                    &             & 12/09/2017 &        & 7$\farcs$8 & 144-154 & 0.99 &4.7.2\\

\hline 
& & & Band 6 & & &  \\ \hline                  
     2016.2.00117.S & Yoshida, K. & 31/08/2017 & 26943  & 4$\farcs$3 &   230-245 & 1.12 & 4.7.2\\
                    &             & 01/09/2017 &        & 4$\farcs$3 &   230-245 & 1.09 & 4.7.2\\
                    &             & 15/09/2017  &        & 4$\farcs$3 & 225-245  & 1.11 & 4.7.2 \\
                    &             & 16/09/2017  &        & 4$\farcs$3 & 225-245  & 1.10 & 4.7.2 \\
                    &             & 16/09/2017  &        & 4$\farcs$3 & 225-245  & 1.02 & 4.7.2 \\
                    &             & 16/09/2017  &        & 4$\farcs$3 & 225-245  & 1.10 & 4.7.2 \\
                    &             & 17/09/2017  &        & 4$\farcs$3 & 225-245  & 1.03 & 4.7.2 \\
                    &             & 17/09/2017  &        & 4$\farcs$3 & 225-245  & 0.99 & 4.7.2 \\      
                    &             & 17/09/2017  &        & 4$\farcs$3 & 225-245  & 0.95 & 4.7.2 \\
     2018.1.01205.L & Yamamoto, S. & 24/10/2018 & 4474 & 5$\farcs$9  & 216-235   & 0.98 & 5.6.1-8 \\
                    &              & 20/10/2018 &     & 4$\farcs$8  & 245-262    & 1.07 & 5.6.1-8 \\

\hline
& & & Band 7 & & &  \\ \hline                  
     2016.2.00117.S & Yoshida, K. & 15/09/2018 & 2782 & 3$\farcs$1 & 352-363 & 1.15 & 4.7.2 \\
\hline      
                                    
\end{tabular}
\end{table*}

\section{ALMA observations}
\label{sec:data_red}
We here summarize the details of the observations used throughout this work. This source has been extensively studied with both the extended and compact arrays of ALMA. Archival data include high-resolution observations designed to detect disk substructure as well as lower-resolution, high-sensitivity observations of the extended envelope. We collect every available ALMA and ACA dataset suitable for our study of L1527, aiming to combine them in the deepest continuum analysis of the envelope so far. A summary of the datasets used in this work can be found in what follows and in Tables \ref{table:1} and \ref{table:2}.

\subsection{ALMA FAUST B3 and B6}
Here, we present new B6 and B3 data from the ALMA Large Program Fifty AU Study of the chemistry in the disk/envelope system of Solar-like protostars (FAUST, PI: Yamamoto, S.). The main goal of FAUST is to investigate the gas chemical composition of the environments surrounding young, Sun-like protostars including their embedding envelopes, their outflows, and disks (see \citealt{Codella2021} for further details).

The observations aimed at L1527 were centered at right ascension $\alpha$(2000) $=$ 04h39m53.878s and declination $\delta$(2000) $= +$26$^{\circ}$03$'$09.56$''$. B3 observations were run on December 14, 2018 and August 25, 2019 with the 12m array in ALMA configurations C43-3 and C43-6, respectively. The baselines of these observations thus ranged 15-2500 meters. ALMA observed L1527 in the 93-95 and 104-108 GHz ranges. The source was observed for approximately 35 minutes in B3, for a continuum sensitivity of 0.025mJy. B6 observations were performed between October 20th, 2018 and December 15, 2019 with both the 7m and 12m arrays. The baselines of the 7m array ranged from 8.9 to 48.9 meters. The 12m array observations were taken in configurations C43-2 and C43-5, with baselines ranging from to 15-314 meters and 15-1400 meters, respectively. ALMA observed in the 244-247 and 258-262 GHz ranges with a spectral resolution of approximately 62.5 MHz. The source was observed for approximately 75 minutes in B6, for a continuum sensitivity of 0.026mJy. 

\begin{figure}[t]
    \centering
    \includegraphics[width=0.66\linewidth]{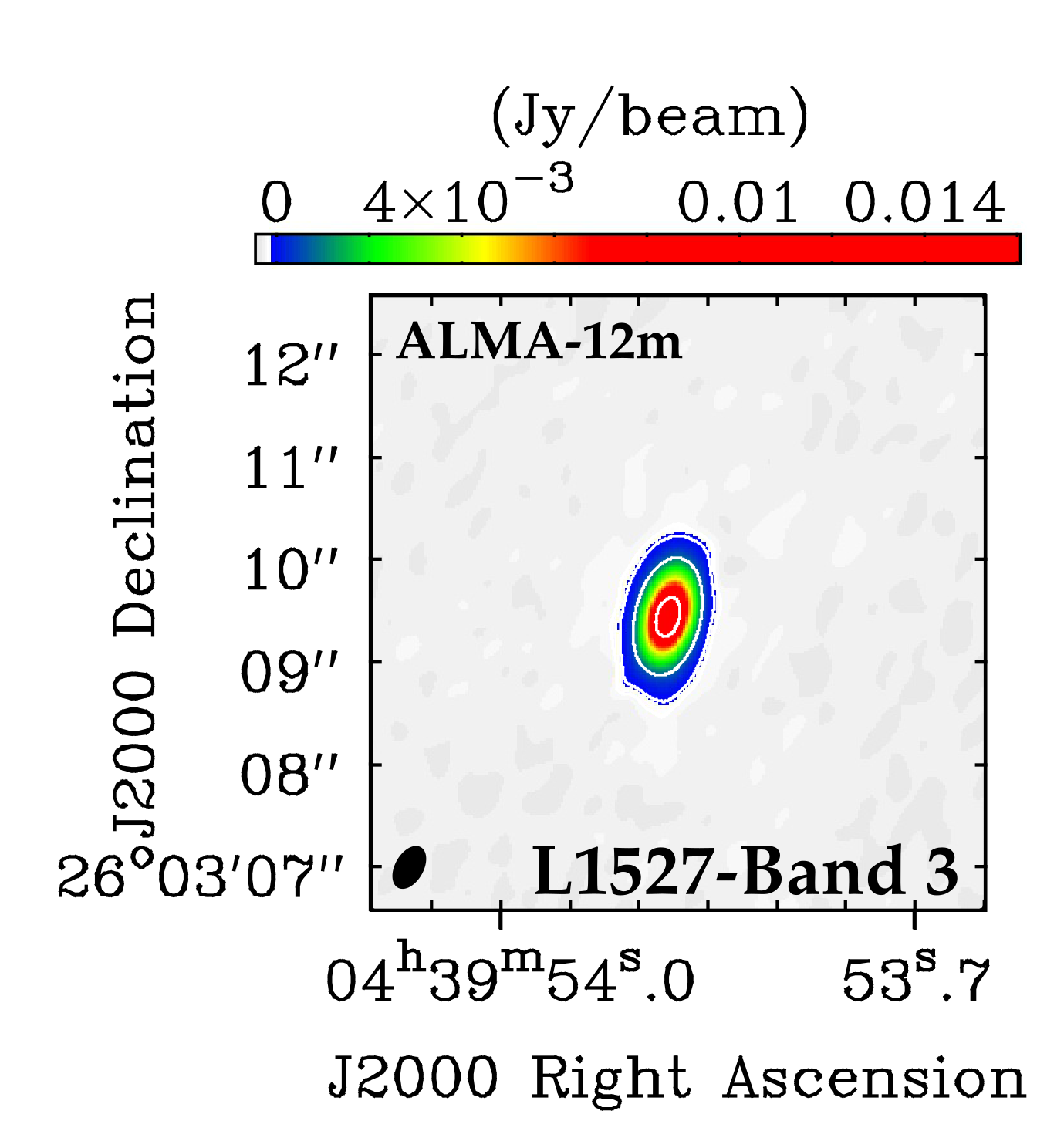}
    \caption{Zoom in to the inner 6$\farcs$0 of the ALMA B3 continuum of L1527 obtained with the 12m array setups of the FAUST Large Program. The color map shows only flux densities higher than $\sim$5$\sigma$, with the white contours highlighting the [5, 15, 30]$\sigma$ levels. The ALMA synthesized beam is shown in black in the lower left corner.}
    \label{fig:faustb3}
\end{figure}

\begin{figure}[t]
    \centering
    \includegraphics[width=\linewidth]{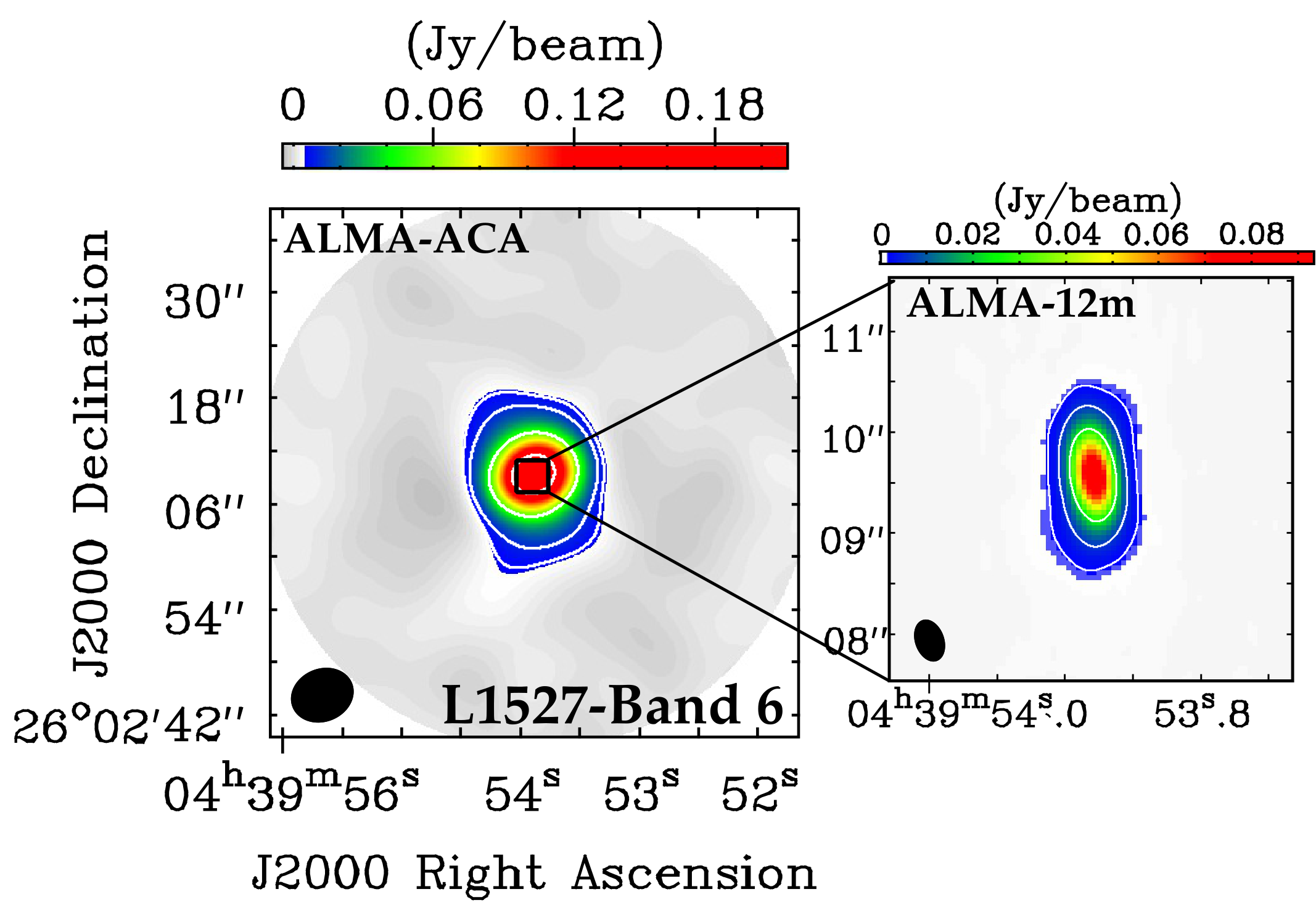}
    \caption{ALMA B6 continuum of map of L1527 obtained imaging the 7m array setup of the FAUST Large Program. The inner 4$\farcs$0 inset has been imaged using ALMA B6 FAUST 12-m array setup data.
    The color map for the ACA image shows only flux densities higher than 5$\sigma$, and the white [5, 10, 50, 150]$\sigma$ level contours.
    The color map of the 12m-array image shows only flux densities higher than 20$\sigma$, and the white [20,60,100]$\sigma$ level contours. 
    The synthesized beams are shown in black in the lower left corner in both cases. We note that the colorbar is different for each map.}
    \label{fig:faustb6}
\end{figure}

The FAUST data were calibrated using a modified version of ALMA pipeline version 42866, using the Common Astronomical Software Applications (CASA, \citealt{McMullin2007}), version 5.6.1-8. This included a correction for errors introduced by the per-channel normalization of data by the ALMA correlator\footnote{https://help.almascience.org/kb/articles/what-errors-could-originate-from-the-correlator-spectral-normalization-and-tsys-calibration}. Line-free Local Standard of Rest Kinematic (LSRK) frequency ranges were identified by visual inspection and averaged per spectral window, and initial continuum images were produced for each separate ALMA configuration. These were then used as initial models for subsequent per-configuration phase-only self-calibration, followed by amplitude and phase self-calibration. Great care was taken to ensure that the models were as complete as possible to avoid changing the overall flux density scale of the data when doing amplitude self-calibration. L1527 is sufficiently strong that per-integration phase-only self-calibration was possible, while for amplitude self-calibration per-scan self-cal was used.
The per-configuration datasets were then aligned across configurations in both phase and amplitude, again using a self-calibration technique. Corrections to the amplitude scale of up to 10\% were found to be needed for some datasets. Improvements in the dynamic range (peak/RMS away from emission) of more than an order of magnitude for the final images were achieved using this technique for setup 1 and 2. The improvement for setup 3 was $\sim35\%$.
Finally, we imaged the visibilities with the CASA \textit{tclean} function, using the hogbom deconvolver, a "briggs" weighting scheme with robust parameter set to 0.5. The resulting beam in B3 is 0$\farcs$44x0$\farcs$27 wide with PA=-29$^{\circ}$; while the synthesized beam in B6 is 0$\farcs$42x0$\farcs$28, PA=19$^{\circ}$. We present the new continuum maps in Fig. \ref{fig:faustb3} and Fig. \ref{fig:faustb6}.

\subsection{ALMA Band 3}
We gathered ALMA Band 3 (B3) observations of L1527 spanning from 2017 to 2020. Given the different scientific aims of the projects, the data collected with the 12-m array has a resolution between 0$\farcs$09 (project 2017.1.00509.S, PI: Sakai, N.) and 2$\farcs$4 (project 2015.1.00261.S, PI: Ceccarelli, C.). The sensitivity of the 12-m array observations that have been used in this work ranges from 0.03 to 0.06 mJy and the total time on source is $\sim$210 minutes. The frequency of the side bands ranges from 85 to 115 GHz.\\
The 7-m array has been pointed at L1527 in B3 for a total $\sim$102 minutes (projects: 2016.1.01541.S and 2018.1.00799.S; PIs: Harsono, D. and Pineda, J.). The observations have resolutions 12$\farcs$0 - 13$\farcs$2 and have been carried out in a range of frequencies from 92 to 105 GHz with sensitivities of 0.3-0.6 mJy.

\subsection{ALMA Band 4}
L1527 has been observed in Band 4 (B4) throughout several months in 2016 and 2018. The data collected with the 12-m array has a resolution between 0$\farcs$1 (project 2016.1.01203.S, PI: Oya, Y.) and 2$\farcs$1 (project 2016.1.01541.S, PI: Harsono, D.). The sensitivity of the 12-m array data that have been used in this work ranges from 0.03 to 0.3 mJy and the total time on source is $\sim$82 minutes. The frequency of the side bands ranges from 138 to 154 GHz.\\
The 7-m array observed L1527 in B4 for a total $\sim$107 minutes (projects: 2016.1.01541.S and 2016.2.00171.S; PI: Harsono, D.). The observations have resolutions 7$\farcs$8 - 9$\farcs$1 and have been carried out in a range of frequencies from 144 to 154 GHz with sensitivities of 0.2-0.4 mJy.

\subsection{ALMA Band 6}
ALMA Band 6 (B6) observations of L1527 span from 2013 and 2020. The data collected with the 12-m array has a resolution between 0$\farcs$2 (projects 2012.1.00193.S, 2013.1.00858.S and 2013.1.01086.S; PIs: Tobin, J.; Sakai, N. and Koyamatsu, S.) and 1$\farcs$1 (project 2018.1.01205.L; PI: Yamamoto, S.). The sensitivity of the 12-m array observations that have been used in this work ranges from 0.03 to 0.09 mJy and the total time on source is $\sim$403 minutes. The frequency of the side bands ranges from 218 to 263 GHz.\\
The 7-m array observed L1527 in B6 for a total $\sim$85 minutes. The observations have resolutions spanning 4$\farcs$3 - 5$\farcs$9 (projects: 2016.2.00117.S and 2018.1.01205.L; PIs: Yoshida, K. and Yamamoto, S.) and have been carried out in a range of frequencies from 216 to 262 GHz with sensitivities of 0.4-1.0 mJy.

\subsection{ALMA Band 7}
L1527 has been observed in Band 7 (B7) from 2014 and 2018. The data collected with the 12-m array has a resolution between 0$\farcs$075 (project: 2016.A.00011.S, PI: Sakai, N.) and 0$\farcs$5 (project 2011.0.00604.S; PI: Sakai, N.). The sensitivity of the 12-m array observations that have been used in this work ranges from 0.06 to 0.2 mJy and the total time on source is $\sim$155 minutes. The frequency of the side bands ranges from 218 to 263 GHz.\\
The 7-m array observed L1527 in B6 for a total $\sim$46 minutes during project 2016.2.00117.S (PI: Yoshida, K.). The observations have resolutions 3$\farcs$1 and have been carried out in a range of frequencies from 352 to 363 GHz with sensitivity of 1.44 mJy.


\subsection{Data reduction and self-calibration}
Since we work with data taken at several frequencies, we avoid frequency smearing by splitting datasets with a relatively large frequency offset. This is especially important in B3. Thus, we split its datasets into two groups.
This way, after calibration, we work with five final frequencies: 88, 100, 141, 249, and 348 GHz. These frequencies roughly correspond to 3.4, 3.0, 2.1, 1.2 and 0.88 mm wavelengths.

The first round of calibrations was performed using the standard CASA pipeline methods provided by the ALMA Regional Centers (ARC). Considering the data were acquired over a period of several years, different versions of CASA were used for these calibrations. The version used for each dataset is listed in Table \ref{table:1}. This first round includes system temperature, phase, amplitude, and bandpass calibration, along with corrections to account for atmospheric water vapor.

We carried out further data reduction and calibration steps using CASA version 6.2.1.7. First, the ALMA cubes for each execution block were inspected and additional flagging was applied when necessary, that is to mask out spectral lines. We are only interested in the continuum emission, thus, to speed up further operations, we channel-averaged the spectral windows of every execution block. We used a common width to level out the S/N among spectral windows and we were careful to avoid bandwidth smearing effects by limiting the averaging based on the maximum baselines and spectral window bandwidth \citep{Bridle1999}. Before combining the datasets for the analysis, we performed several additional operations. First, we ran the \textit{statwt} CASA routine in order to fix the visibility weights. This is especially important for datasets from old ALMA Cycles that have been reduced with CASA 4.2.1 or earlier versions\footnote{https://casaguides.nrao.edu/index.php/DataWeightsAndCombination}.
We then imaged the averaged-channel visibilities with the \textit{tclean} task and fitted a Gaussian with \textit{imfit} to pinpoint the maximum of the emission. Next, we used the \textit{fixvis} function to shift the phase center to the position of the peak of L1527 at the epoch of every observation. 
\begin{figure*}[ht]
    \centering
    \includegraphics[width=\textwidth]{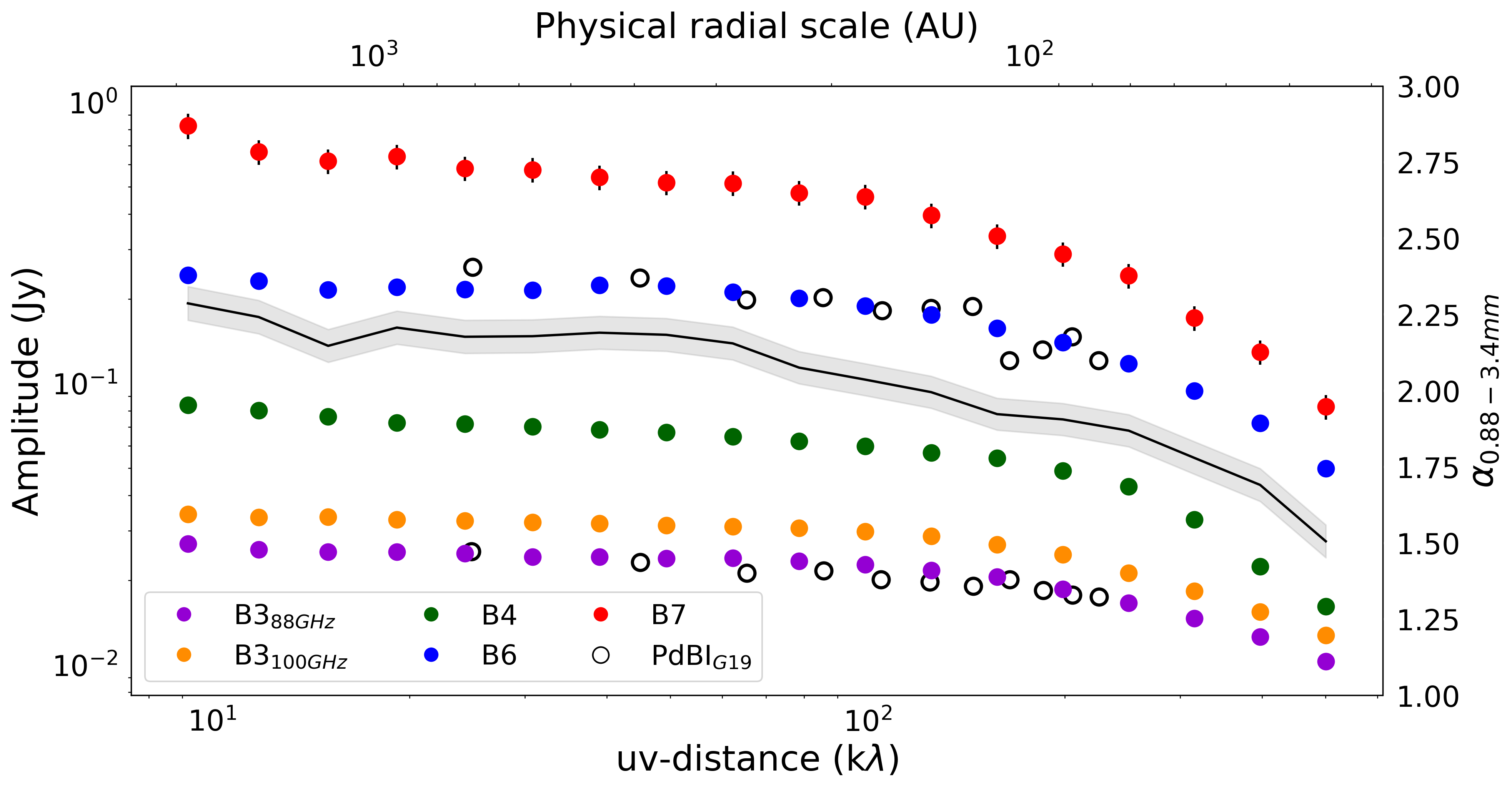}
    \caption{L1527 ALMA  88 GHz (purple), 100 GHz (orange), 141 GHz (green), 249 GHz (blue), 348GHz (red) visibility amplitudes up to 1000 k$\lambda$. 
    We also compare the 94 and 231 GHz observations (white points) performed with the Plateau de Bure Interferometer (PdBI) and anlayzed in \citet{Galametz2019}.
    The slope of our observed Spectral Energy Distribution (black line) has been obtained by fitting a line to the fluxes at all wavelengths in each uv-distance bin. While $1.5 < \alpha_{fit} < 2.5$, most of the contribution to the flux - and to $\alpha$ - at any uv-distance is due to the disk (see Section \ref{section:spindx}).}
    \label{fig:our_data}
\end{figure*}
Since the target coordinates were slighlty different across datasets, we also set the source sky position (in the metadata) to a common center across execution blocks using \textit{fixplanets}. Once the coordinates of each dataset were phase-centered and aligned following this procedure, we rescaled the fluxes following the procedure of \citet{Andrews2018}: for every frequency, selected a range of uv-distances where the uv coverage of two datasets overlaps, binned the visibilities in that range, and compared them to obtain a rescaling factor where the visibilities overlapped. The rescaling factors were in a 1-15\% range for all execution blocks. The rescaling correction was then computed using the task \textit{gencal} and the rescaling itself was achieved by applying the correction factors with \textit{applycal}. We report the rescaling factors in Tables \ref{table:1} and \ref{table:2}.

Based on this model, we then computed the phase corrections on each execution block using \textit{gaincal} and applied them with \textit{applycal}. This step aimed to correct phase errors between executions and between spectral windows. We repeated this procedure three times. In the first round, we set the solution interval of \textit{gaincal} to \textit{inf}, combined the scans within each execution block and set gaintype $=$ "G", that is the gains were determined for each polarization and spectral window. In the second and third rounds, we combined the spectral windows of each block, shortened the solution intervals to 60 and 10 seconds, respectively, and set gaintype $=$ "T", to obtain one solution for both polarizations. The self-calibration yielded peak S/N improvements in the range 10-500 \% when evaluated on the images obtained with the combined datasets. We did not find any appreciable improvement in the noise and signal-to-noise properties for phase-only self-calibration steps with even smaller time intervals. At this point, we combined the corrected datasets using the \textit{concat} CASA task. 

Since we analyze extended emission, we estimated the effect of the antennas primary beam (PB) correction on the recovered flux. In B7, the ALMA 7m and 12m antennas PB full width at half maximum is 33$\farcs$0 and 19$\farcs$0, respectively. This results in a loss of about 10\% of flux at 20 k$\lambda$ (1000 au scales), for the 12m antennas. This only marginally affects the recovered spectral index we are interested in, increasing it by a factor of $<$5\% even at the largest scales. 

\subsection{L1527 as seen in the visibility space}
\label{literature_comparison}

The final binned visibility amplitudes obtained with our data reduction and self-calibration are illustrated in Fig. \ref{fig:our_data}. We also show that the associated spectral index lies in the range $1.5 < \alpha_{0.88-3.4mm} < 2.4$. This low $\alpha$ is not surprising as the disk emission contribution dominates at every uv-distance (see Section \ref{sec:model}), thus the total spectral index is strongly influenced by the optically thick disk (cfr. Section \ref{section:spindx}).  

Among other class 0/Is of the CALYPSO survey, the continuum emission of the envelope infalling on L1527 has been studied in \citet{Galametz2019}, hereafter G19, who used PdBI observations at 1.3 and 3.2 millimeter. The data they used spans a range in uv-distances from 20 to 200 k$\lambda$ at both frequencies.  Fig. \ref{fig:our_data} shows a comparison between the ALMA observations we used and G19's PdBI fluxes (white points). 

We cut the ALMA datasets at the shortest common baseline, set by B7 at 9.8 k$\lambda$, hence a Maximum Recoverable Scale (MRS) of roughly 26$\farcs$0, or 3500 au at the distance of the source.
We aim to more robustly determine the spectral and dust-emissivity indices of the source by extending the frequency range, that dominates the error on $\alpha$ (see Section \ref{section:spindx}). Furthermore, a denser sampling of frequencies is critical to understand the optical depth of the source across wavelengths, which is vital to constrain where the spectral index is a good proxy for dust grain sizes.
On top of enriching the studied frequency range, we used a combination of ALMA and ACA data to study the dust continuum emission with high S/N at scales 2 times larger than G19, toward both shorter and longer baselines. In fact, although our primary attentions are devoted to the envelope, long-baseline visibilities are of great importance if we want to properly model the compact emission. G19 subtracted a constant amplitude (F$_{200 k\lambda}$) from each visibility at the two frequencies, assuming an unresolved disk that would contribute a constant flux at baselines shorter than 200 k$\lambda$ ($\sim$1$\farcs$0, or 140 au). 

Considering the wider uv-coverage of the data we used (9.8-1000 k$\lambda$), we can more precisely model and subtract the compact emission by fitting a Gaussian component within a model that separately accounts for envelope and disk (see Section \ref{sec:model}). This procedure also makes a significant difference when computing the errors on $\alpha$ and $\beta$, which in turn depend on the relative errors of the amplitudes. Since the PdBI amplitudes in G19 were roughly constant and only slightly above F$_{200 k\lambda}$, the relative error on the fluxes after the subtraction of F$_{200 k\lambda}$ became large already at a uv-distance of $\sim$ 100 k$\lambda$. Given the high S/N of our combined datasets, the relative flux uncertainty remains relatively contained after disk subtraction (see Fig. \ref{fig:mean_alpha}). In the upcoming Section, we lay out the methods we followed to model the full visibilities and subtract the disk component to obtain an envelope-only spectral index.

\section{Disentangling disk and envelope emission}
\label{sec:model}
To investigate dust properties in the envelope of L1527, we worked in the visibility ($u, v$) plane. This approach has the main advantage of keeping the analysis clear of the artifacts and non-linearity from which image reconstruction algorithms suffer.

The visibility amplitudes are the combined flux of disk and envelope at all $uv$ distances. Since we are mainly interested in studying the spectral index of the envelope, we have to disentangle the compact emission flux from the extended one. Hence, we need to fit the data with a model appropriate for a compact disk that, at the same time, self-consistently accounts for excess emission at the shortest baselines (largest physical scales).

Brightness profiles of protostellar envelopes of diverse class 0/I YSOs have been found to be well fit by broken power laws with exponents in the (-2.5,-1) range (\citealt{Plummer1911}, \citealt{Motte2001}, \citealt{shirley2002}, \citealt{Maury2019}). Thus, we consider a geometrical model in which a Gaussian, that will trace the compact emission, is summed to a power law (Plummer profile):
\begin{equation}
    I_{TOT}(R) = rI_0 \cdot e^{-\frac{R^2}{2\sigma^2}}  + \frac{(1-r) I_0}{\Bigg(1 + \Big(\frac{R'}{R_i}\Big)^2 \Bigg)^{\frac{p-1}{2}}} .
    \label{eq:total_profile}
\end{equation}

\begin{figure}[t]
    \centering
    \includegraphics[width=\linewidth]{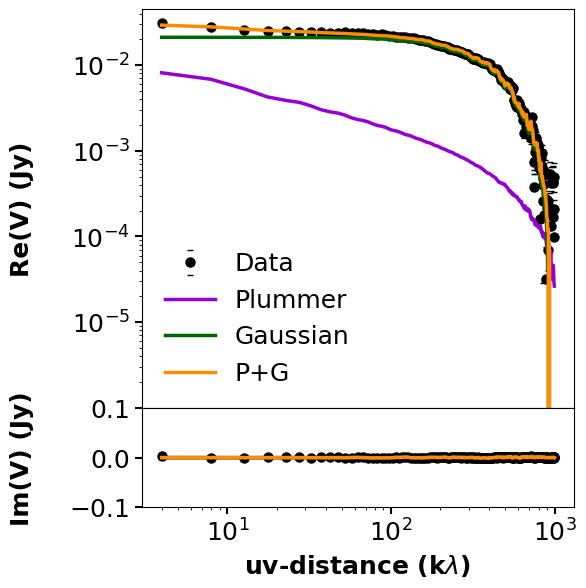}
    \caption{Final Plummer+Gaussian best fit (orange) is overplotted on the real part of the visibilities for the B3 (88 GHz) data (upper panel, black points). The fit on the imaginary is in the lower panel.
    The Plummer only (violet line) and Gaussian only (green line) components of the total model are also shown. In the case of the 88 GHz emission, the disk contributes up to 98\% of the total emission. The wiggles in the model are due to its sampling on the uv points of the observations.}
    \label{fig:fitband3}
\end{figure}
The Gaussian is defined by a peak $f_0' = rI_0$, a width $\sigma$, an inclination $inc$ and position angle PA. The envelope power law has peak flux $f_0'' = (1-r)I_0$ and three free parameters: $R_i$, $R_{out}$, $p$. These are such that the envelope emission is constant within a radius $R_i$, it follows Eq. \ref{eq:total_profile} between $R_i$ and $R_{out}$ and it is null beyond $R_{out}$. 
The peak fluxes of the extended and compact emission are modulated by the free parameter $0 < r < 1$.
We have performed the fitting with \texttt{galario}, a library that exploits the power of modern graphical processing units (GPUs) to accelerate the analysis of observations from radio interferometers \citep{Tazzari2018}. In our framework, \texttt{galario} would compute a model image given our total profile, then it would Fourier-transform it into synthetic visibilities and sample them at the uv-points covered by the antenna configurations with which the observations were performed. Then, it would run a minimum-$\chi^2$ fit between the data and the model visibilities. 
The 10 fit parameters were the flux amplitude $I_0$ and the $r$ factor; the width $\sigma$, the inclination (inc) and position angle (PA) of the Gaussian; the inner and outer envelope radii, $R_{in}$ and $R_{out}$, and the power law exponent $p$. Finally two parameters, dRA and dDec, fit the offset of the peak from the phase center. For each fit, 80 walkers were set to run for 3,000 burn-in steps and 7,000 more steps after the burn-in stage.
The best fit results are summarized in Table \ref{tab:fits_results}. We show, as an example, the single best-fit model for the B3 88 GHz data, decomposed in its power law and Gaussian components (Fig. \ref{fig:fitband3}). We report the rest of the fits in Figures \ref{fig:B3fit}, \ref{fig:B4fit}, \ref{fig:B6fit}, \ref{fig:B7fit}, along with the best fit parameters summary (Table \ref{tab:fits_results}) in Appendix \ref{sec:appendix2}.
In particular, we find that the compact emission has a deprojected width of $0\farcs15 < \sigma_{fit} < 0\farcs28$ on the sky, depending on the Band. If we define the radius of the disk to be the $2\sigma$ contour of the Gaussian component, then $R_{disk} \sim 75$ au in B7, consistent with what was found in the kinematical analysis of \citet{Aso2017}. In addition, the Gaussian component contributes to a minimum of 75\% in B4, up to 98\% in B3 (see Appendix \ref{sec:appendix2} for the detail). The derived disk fluxes are listed in Table \ref{tab:disk_tab}. This fitting procedure allows us to disentangle disk and envelope emission and study the spectral energy distribution of the disk and envelope separately in the next section.

\begin{figure}[t]
    \centering
    \includegraphics[width=\linewidth]{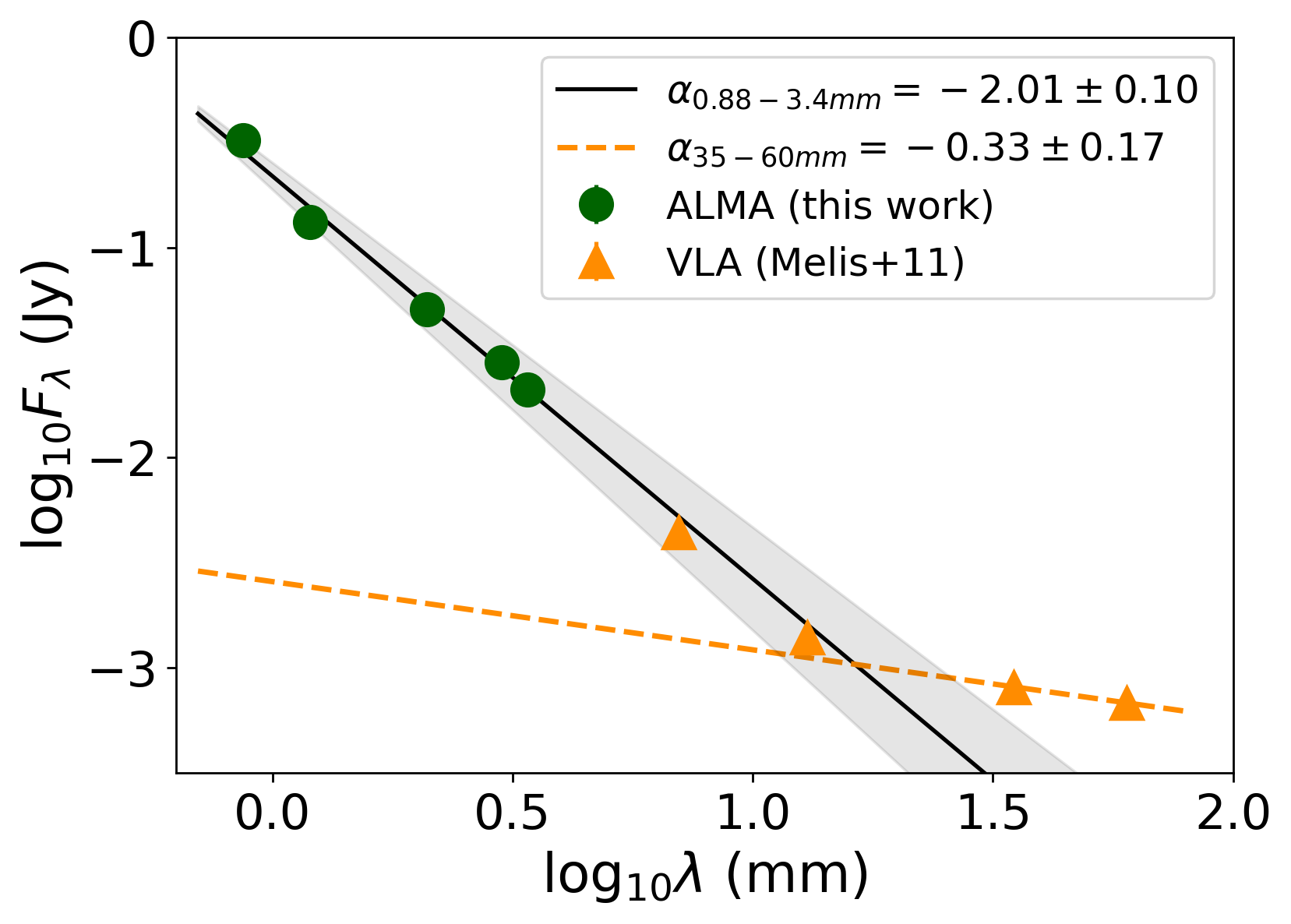}
    \caption{Spectral energy distribution of the circumstellar disk around L1527. Our \textsf{galario} best-fit total flux of the compact Gaussian component for each band, in green, along with longer wavelength VLA measurements from \citet{Melis2011} (see Table \ref{tab:disk_tab}). 
    The protoplanetary disk of L1527 is optically thick up to 3.4 mm ($\alpha_{B7-B3} \sim$ 2), likely due to its edge-on nature.}
    \label{fig:disk_sed}
\end{figure}

\section{Spectral and dust emissivity indices}
\label{section:spindx}


In this Section, we first derive the spectral indices of the observed emission at both disk and envelope scales, then we discuss to what extent these can be used to evaluate dust grain properties.
\begin{figure*}[t]
    \centering
    \includegraphics[width=\textwidth]{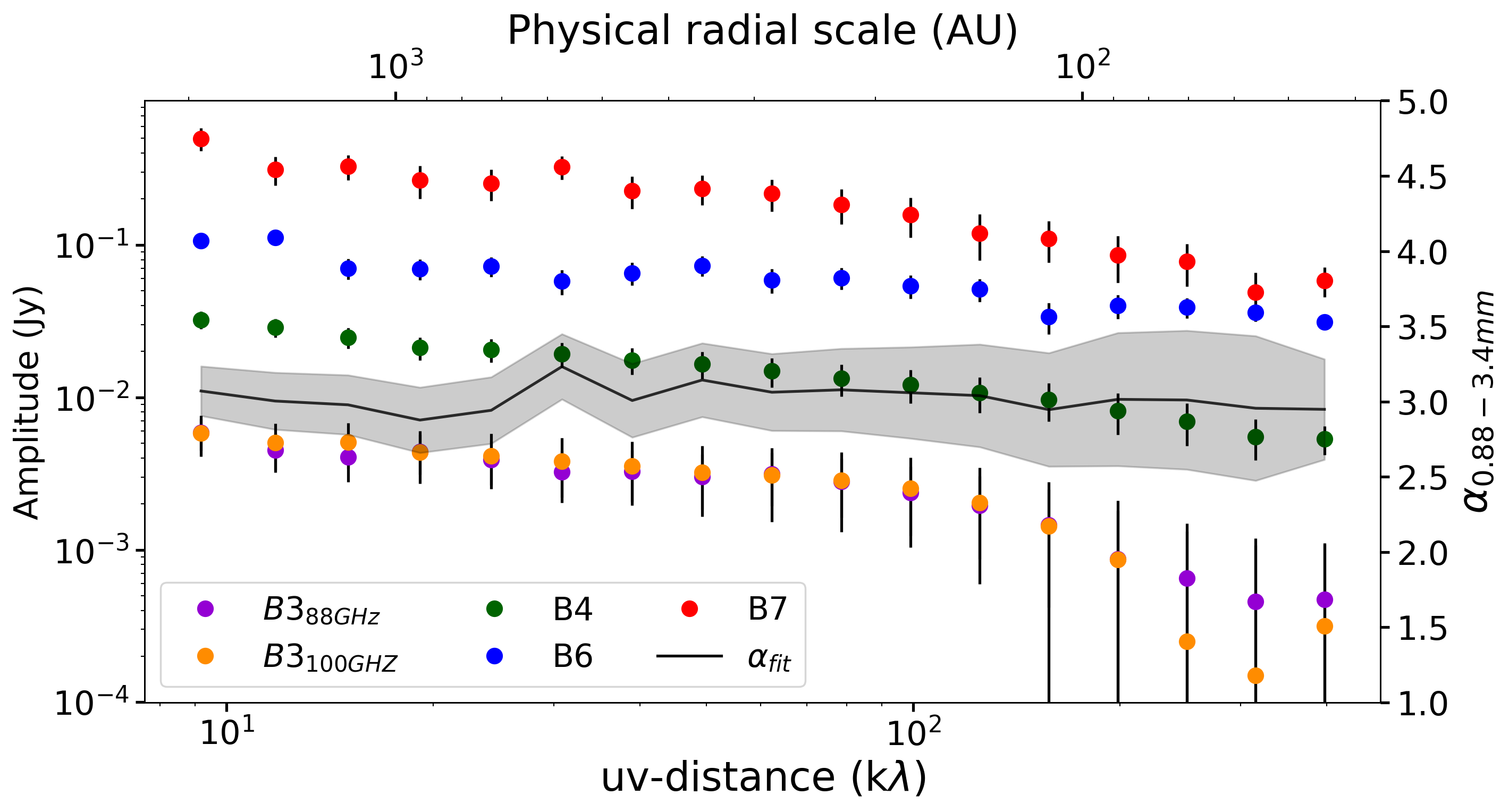}
    \caption{L1527 ALMA B3 (orange), 4 (green), 6 (blue), 7 (red) visibilities after removal of the fitted compact Gaussian component. The spectral index of the envelope emission (gray) has been obtained by fitting a line to the fluxes at all wavelengths in each uv-distance bin.}
    \label{fig:mean_alpha}
\end{figure*}
\subsection{The circumstellar disk}
\label{sec:spindx_disk}
To study possible early dust grain growth in the envelope of L1527, we made use of long baseline data to properly model the compact emission to be subtracted from the short baseline amplitudes. In doing so, we obtained estimates of the flux of the compact ($0.15" < \sigma_{fit} < 0.27"$) emission of L1527 in four ALMA bands. 
We here show the spectral energy distribution of the disk we obtained by combining and modeling all suitable ALMA archival data for L1527 (Figure \ref{fig:disk_sed}, Tab. \ref{tab:disk_tab}). To offer a more comprehensive view to the reader, we extended the range of wavelengths by including literature Very Large Array (VLA) measurements from \citet{Melis2011}, up to 6 cm. 

Fitting the SED in the 0.88 to 7-millimeter range using our measurements along with the values of \citet{Melis2011}, we find an optically thick disk with $\alpha = 2.1 \pm 0.1$. This is not surprising: the disk orbiting L1527 is very inclined ($i \sim 80^{\circ}$), almost edge-on to the sky: this geometric factor contributes to increase its optical depth along our line of sight. Furthermore, \citet{Ohashi2022} used high-resolution, multiple frequency observations of the disk and found that the brightness temperature toward the mid-plane of the disk obtained by converting the flux at 0.88 mm (42 K) is much lower than that found at 2 and 3 mm (60 K and 90 K, respectively). This progressive increase in brightness temperature with wavelength means that the disk is at least partially optically thick at 0.88 and 2 mm. 
 
The high optical depth of the circumstellar disk of L1527 prevents us from concluding anything about its dust grain properties based on the spectral index alone. 
Finally, the 3.5 and 6 cm fluxes can be used to estimate the free-free emission that affects the measurements at the shorter studied wavelengths.
\begin{table}[ht]
\caption{L1527 disk SED built with ALMA (this work) and VLA \citep{Melis2011} observations.}
\label{tab:disk_tab}
\centering
\begin{tabular}{lccc}
\hline 
    $\nu$ (GHz) & $\lambda$ (mm) & Flux density (mJy) & Reference  \\ \\ \hline
     348 & 0.86 & 324 $\pm$ 29 & this work\\
     249 & 1.2 & 132 $\pm$ 9 & this work\\
     141 & 2.1 & 52 $\pm$ 3 & this work\\
     101 & 3.0 & 28 $\pm$ 0.6 & this work\\
     88  & 3.4 & 21 $\pm$ 0.5 & this work\\
     43.5 & 7.0 & 4.4 $\pm$ 0.6 & \citet{Melis2011} \\
     22.5 & 13.3 & 1.4 $\pm$ 0.1& \citet{Melis2011} \\
     8.5 & 35.2 & 0.81 $\pm$ 0.03& \citet{Melis2011} \\
     5 & 60 &0.68 $\pm$ 0.04 & \citet{Melis2011} \\
\hline
\end{tabular}
\end{table}
The slope at these longer wavelengths is $\alpha_{35-60} = 0.33 \pm 0.17$, as reported in Fig. \ref{fig:disk_sed} and is compatible with what is expected for free-free emission \citep{Panagia1975}.
However, it is worth noting that \citet{Melis2011} showed how the 3.5 cm flux of L1527 is highly variable in time. Indeed, they listed eight measurements (references therein) at this wavelength taken between 1996 and 2010 where the 3.5 cm flux shows significant variations between 0.55 and 0.81 mJy, with errors of 1-5\%. In Fig. \ref{fig:disk_sed}, we show \citet{Melis2011} points taken at roughly simultaneous epochs (2010 July 30th). The contribution of the free-free emission is lower than 10\% in B3 and lower than 1\% in B7.

\subsection{The envelope}
\label{sec:spindx_envelope}
We investigate changes in the spectral index throughout the spatial scales of the envelope in the visibility plane. Starting from the common shortest baseline of $\sim$ 9 k$\lambda$, we log-uniformly bin the visibilities across the available uv-distances and then measure the spectral index in each bin. Since we are interested in studying the spectral index of the envelope, we have subtracted the \textsf{galario} best-fit Gaussian component of our model (see Section \ref{sec:model}) from the data. This way, the remaining flux is the contribution of the extended emission only (Fig. \ref{fig:mean_alpha}). 

Here, the amplitude and related error bars account for both statistical and calibration uncertainties. We set the latter to 10\% for B7 and 5\% for bands 3, 4 and 6, following the prescriptions of the ALMA Handbook and on-sky analysis \citep[respectively;][]{Remijan2019,francis20}. While the error on $\alpha$ is dominated by flux calibration errors on the observed amplitudes (see Fig. \ref{fig:our_data}), the relative amplitude error becomes important after the subtraction of the disk model. Not only is the relative error overall larger, it also visibly increases with uv-distance. This effect is clear in Fig. \ref{fig:mean_alpha}. Using almost every archival dataset for this source, we made an effort to maximize the S/N of the faint envelope of L1527 in what is its deepest continuum analysis. 

To make the best use of our multiwavelength data, we fitted a line to the fluxes along all wavelengths in each uv-distance bin and we more robustly determined the spectral index. Since we combined a number of datasets for each band, the frequency of each flux point in each bin is taken to be the weighted average of the frequencies in that bin. The uncertainty on the spectral index obtained by fitting all bands is the error on the slope obtained with the weighted linear regression.
Figure \ref{fig:mean_alpha} shows that the fitted spectral index of the envelope of L1527 appears roughly flat in the studied range that spans from $\sim 2000$ down to 50 au,  $\alpha \sim 3$. For completeness, we show the spectral indices computed between adjacent bands in Appendix \ref{sec:appendix1}. These are flat as well, although they show systematic differences in their average value, which clearly shows the need for a multifrequency (>2) approach.

\begin{figure}[t]
    \centering
    \includegraphics[width=0.82\linewidth]{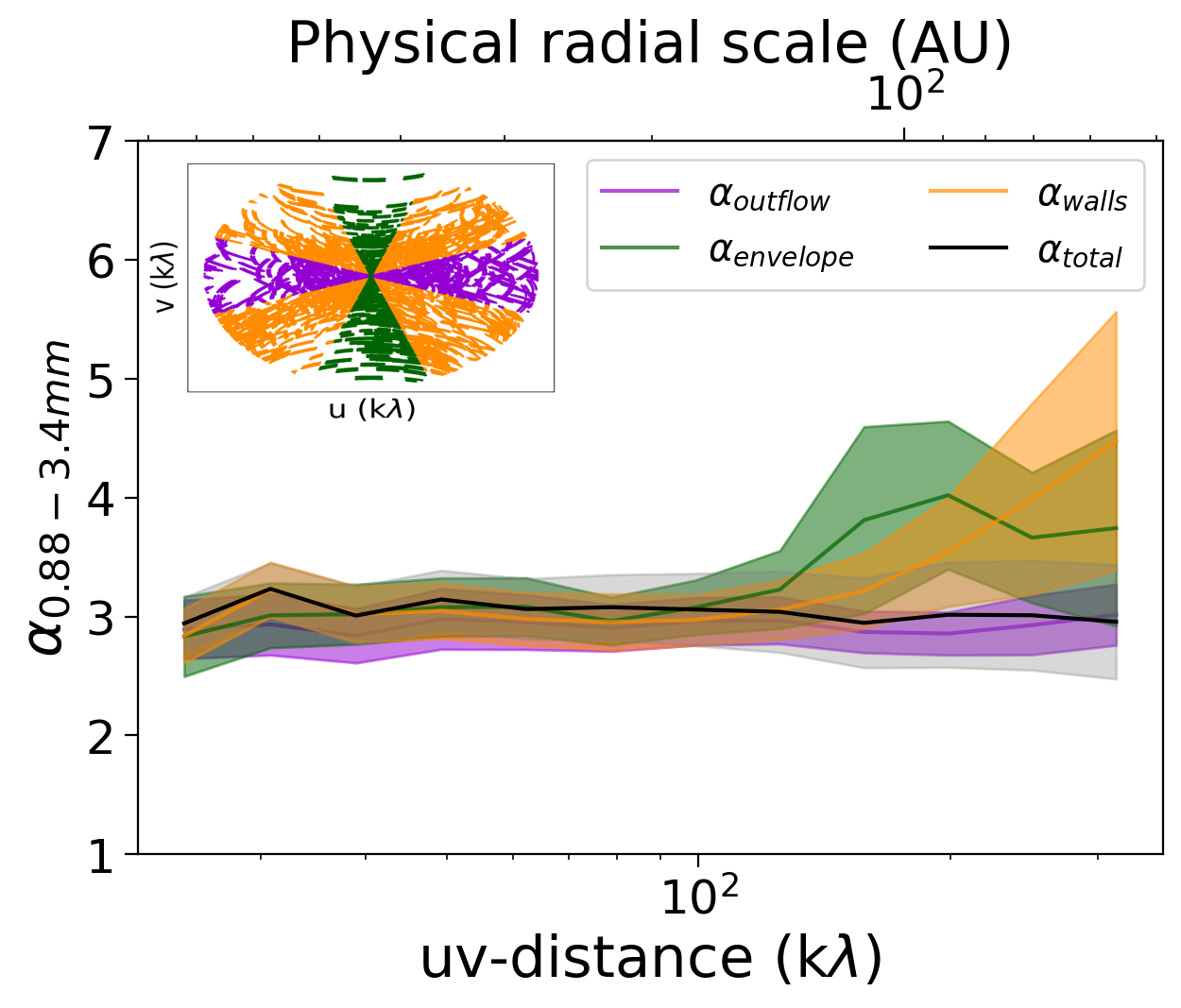}
    \caption{Spectral index of L1527 as measured along directions of the outflow (violet) and envelope (green), cavity walls (orange). The total $\alpha$ is also plotted (gray). The way the uv-coverage has been sampled is shown in the upper inset (B7 as an example). The scarce uv-coverage of the ACA B7 observations cause some bins to be undefined at the short baselines, hence the gaps. In all cases, $\alpha$ is only computed starting from 20 k$\lambda$.}
    \label{fig:spindx_angles}
\end{figure}

Finally, we measured the spectral index in different angle bins on the sky, after removal of the compact component. We divided the uv-plane into three different regions: the envelope ($-20^\circ< PA <20^\circ$), the outflow ($70^\circ< PA <110^\circ$) and the cavity walls (the remaining zones).
We do not observe any significant difference among the three spectral indices at scales larger than 200 au (Fig. \ref{fig:spindx_angles}).

\subsection{Dust emissivity index}
\label{sec:beta_section}
The simple link between the exponent of the dust opacity power law and the spectral index, $\beta = \alpha + 2$, only holds for optically thin emission when the RJ approximation is valid. Thus, before interpreting the values of the envelope's spectral index in terms of dust properties, we check whether these necessary conditions are met at the wavelengths we probe. 
First, we evaluate the optical thickness of the emission. The specific intensity $I_{\lambda}$, or flux per unit solid angle that we receive from the source ($F_{\lambda}/\Omega$) can be generally expressed as an absorbed black body:
\begin{equation}
    I_{\lambda} = \frac{F_{\lambda}}{\Omega} = \big(1 - e^{-\tau_{\lambda}}\big) \cdot B(T_{dust})
\end{equation}
where the optical depth $\tau_{\lambda}$ modulates the difference between the observed flux and the optically thick black body emission. If $\tau_{\lambda} \ll 1$, then $F_{\lambda}/\Omega \sim \tau_{\lambda} B(T_{dust})$, while if $\tau_{\lambda} \gg 1$, then the observed emission tends to a black body spectrum. 

\begin{figure}[t]
    \centering
    \includegraphics[width=\linewidth]{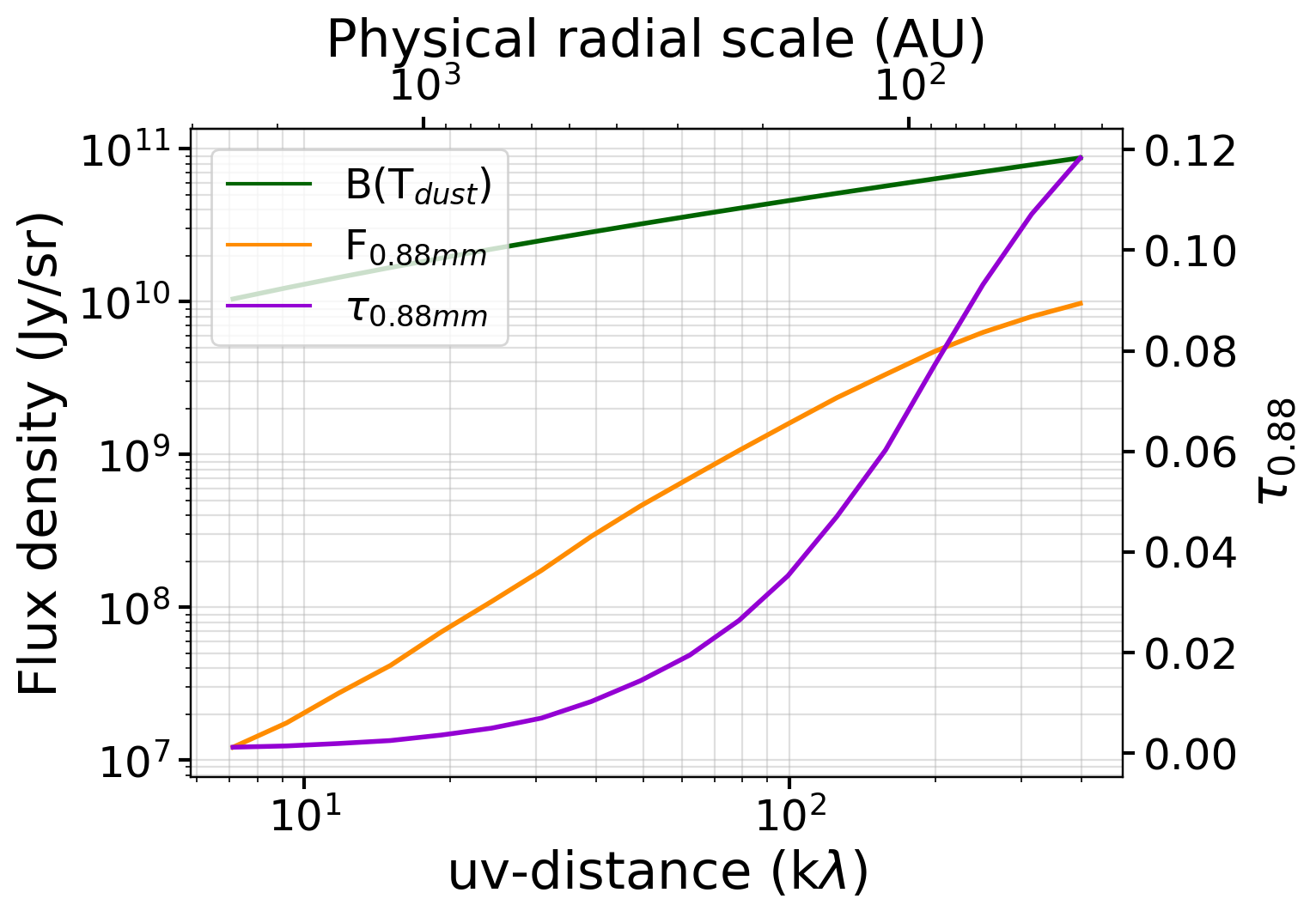}
    \caption{Flux density of the envelope emission in ALMA B7 (orange) is consistently lower by at least an order of magnitude than a black body spectrum of emitting dust with a radial temperature profile $T\propto r^{-0.4}$ (green). The optical depth (violet) is much smaller than 1 at all envelope scales, thus we can consider the envelope emission at 0.88 mm as optically thin.}
    \label{fig:opt_thin}
\end{figure}

To evaluate the optical thickness, we consider a typical dust temperature prescription for a centrally illuminated protostellar envelope (\citealt{Shu1977}, \citealt{Butner1990}, \citealt{Terebey1993}, \citealt{Motte2001}):
\begin{equation}
    T(R) = 38 L^{0.2} \Bigg(\frac{R}{100 au}\Bigg)^{-0.4}.
    \label{eq:temp}
\end{equation}
Considering this relation and the range of scales that we probe, the temperature gets as low as 10 K in the outermost radii of the envelope ($\sim 2000$\,au). We find that, even at the shortest of our wavelengths, the envelope emission is optically thin at all scales as $F_{0.88mm}/\Omega < B(T_{dust})$ by a factor of 10 (Fig. \ref{fig:opt_thin}). The optically thin regime is naturally satisfied at longer wavelengths. 

Secondly, we check if the RJ approximation is valid. Based on the same temperature profile for the dust, in the cold outskirts of the envelope ($\sim$ 20 K at 500 au), $h\nu / k_BT \sim 0.8$ at our mean B7 frequency of 348 GHz. This violates the RJ condition $h\nu / k_BT \ll 1$. Thus, while the envelope of L1527 is in the optically thin emission regime, it does not satisfy the RJ approximation and $\beta \neq \alpha - 2$.

\begin{figure*}[t]
    \centering
    \includegraphics[width=0.8\textwidth]{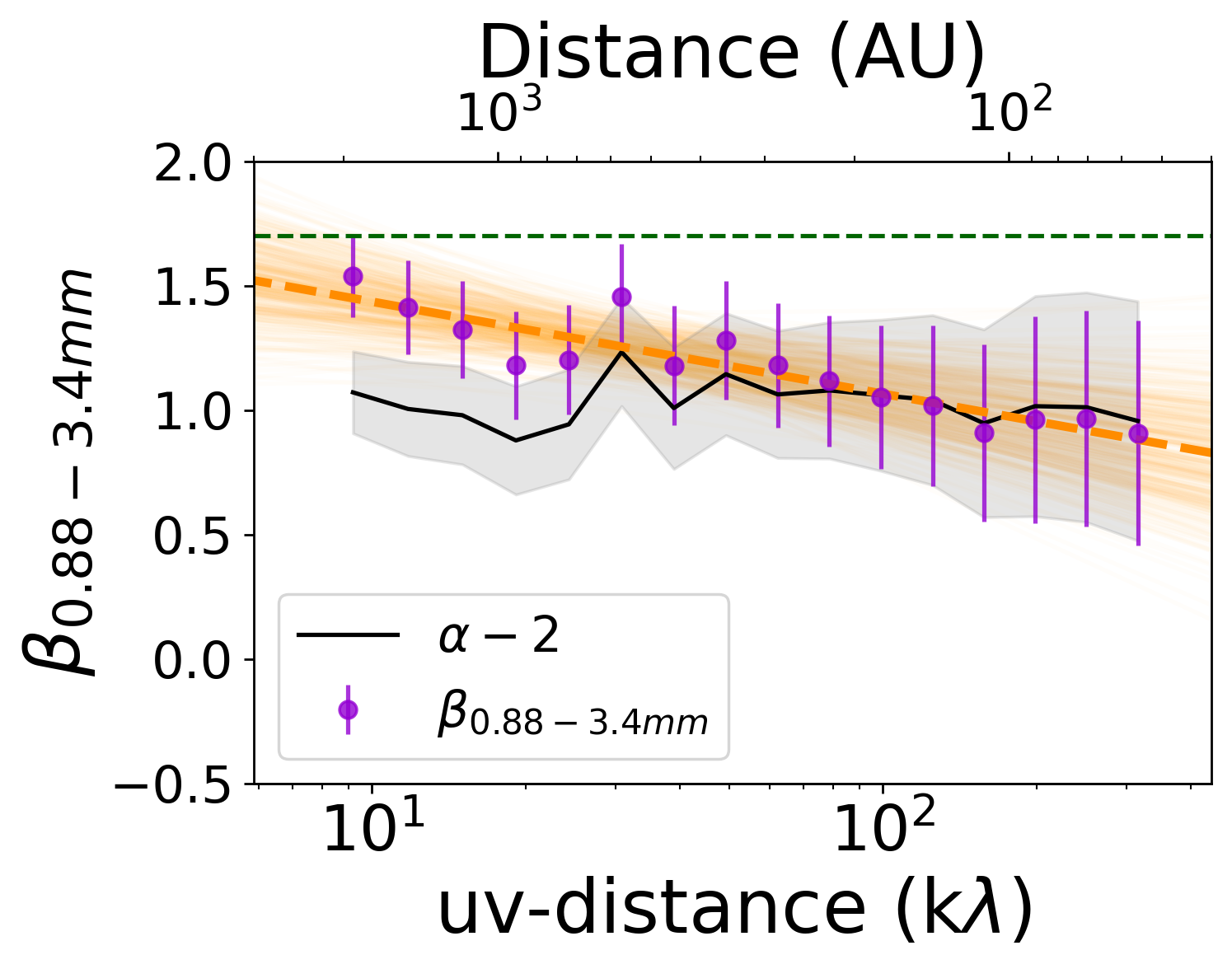}
    \caption{Dust emissivity index (purple dots) of the envelope of L1527 as a function of radial scales. The solid dots indicate the $\beta$ computed fitting all available bands. The dashed orange line is the best model for the profile of $\beta$ along scales, while the light orange lines are a subsample of the results of some chains of the fitting procedure. 
    The black line shows the approximation in which $\beta = \alpha_{fit} - 2$, where $\alpha$ is the slope of the SED considering all available bands.}
    \label{fig:beta}
\end{figure*}

Thus, using the same profile, we can obtain a temperature-corrected $\beta$ profile as:
\begin{equation}
    \log_{10} \Bigg[\frac{F_{\nu}}{B_{\nu}(T)}\Bigg] = \beta \cdot \log_{10}({\nu}) + A ,
    \label{eq:beta}
\end{equation}
where $A$ is a constant (similarly to what done in G19, but for multiple frequencies).
Figure \ref{fig:beta} shows how the dust emissivity index is nearly $\alpha-2$ for the smaller spatial scales, where T(R) is large enough, while it starts to show diskrepancies after 200 au (where T$\sim$ 25 K). We also investigated the effects on $\beta$ of changing the temperature profile's radial power law exponent. We found that exponents in the interval [-0.3,-0.6] do not affect our conclusions. We exclude steeper profiles as they lead to unreasonably low values of the temperature at the studied scales.
The error on the dust emissivity index has been considered to be the same one that affects the spectral index since we are assuming an exact temperature profile, thus $\Delta\alpha = \Delta\beta$. 

In Figure \ref{fig:beta}, we report the measured $\beta$ as a function of uv-distance, and we overplot a linear fit to the points. We have run the fit by means of the Bayesian method described in \citet{Kelly2007} and implemented in \textsf{linmix}\footnote{https://github.com/jmeyers314/linmix}. The priors were uniformly distributed. We plot some single chain results in Fig. \ref{fig:beta}, along with the best fit model. We find evidence for an outward increasing $\beta$, rising from 0.8 to 1.6 between 50 and 2000 au radial scales:
\begin{equation}
    \beta(R_{au}) = 0.37^{+0.17}_{-0.16} \cdot \log_{10}( R_{au}) - 0.23^{+0.45}_{-0.46}.
    \label{eq:beta_grad}
\end{equation}
In turn, Eq. \ref{eq:beta_grad} yields:
\begin{equation}
    \beta(R=300 au) = 1.15 \pm 0.08.
\end{equation}
These measurements represent tentative evidence for dust optical properties variations throughout the envelope of L1527. Furthermore, low $\beta$ at scales lower than $\sim$ 300 au are consistent with the presence of submillimeter sized grains in the inner envelope.

\subsection{Literature comparison}
G19 studied the envelope continuum emission of a few class 0/I sources, including L1527. Here, we briefly compare our findings to theirs. 
First, the wide uv-distance range of the datasets we used allowed us to model the disk emission spatial profile, while only a constant amplitude  was subtracted in the amplitudes in G19, who assumed an unresolved compact disk. This spatial modeling is important to properly subtract the compact emission from the total amplitudes and work out the rest of the analysis on the flux of the envelope alone.

G19 reported a slope of $\beta$ of 0.18 $\pm$ 0.39. Noticeably, this flat $\beta$ profile of G19 is consistent with what we observe across the same region they studied, 70-700 au radii. Extending the studied scale by a factor 2, we are able to explore the outer envelope and determine the positive outward gradient of Eq. \ref{eq:beta_grad}. 
At the outer probed radii ($\sim$ 2000 au), the derived values for $\beta$ are consistent with typical ISM measurements.
Secondly, G19 reported a fitted $\beta_{500 au} = 1.41 \pm 0.16$ while we find $\beta_{500 au} = 1.15 \pm 0.08$ based on Eq.\ref{eq:beta}. These improvements have been obtained thanks to higher sensitivity data as well as the extension of the studied frequency ranges.

\section{Discussion}
We here discuss our results in further detail, how they compare with existing theoretical and observational literature, and the caveats relevant to our analysis. 

\subsection{Dust growth in envelopes is challenging}
\label{section:beta_discuss}
The low $\beta$ value we measured at scales of a few hundred au from L1527 is a tentative evidence for grown dust grains in the inner envelope of the class 0/I YSOs L1527. If solids indeed grew in the envelope, this would represent a shift in the usually assumed initial conditions for planet formation, which only include $< 0.25 \mu$m grains at such scales.
However, a challenge to our measurements rises when trying to link these low $\beta$ with the in situ formation of relatively large dust grains in the envelope environment. 
Indeed, theoretical studies of grain agglomeration have not been able to reproduce such growth in envelope-like environments. 
The numerical simulations of \citet{Ormel2009}, \citet{Lebreuilly2023} and \citet{Bate2022} have shown that it is very unlikely that solid dust grains could grow larger than about 1 $\mu$m at the typical densities and timescales of envelopes such as the one infalling on L1527 and its young disk. These works included a thorough treatment of grain-grain collision energetics, accounting for relative grain velocities due to turbulence, Brownian motions, ambipolar diffusion and hydro-dynamical drift.
Additionally, \citet{Silsbee2022} proposed a simpler, analytical model to place strict upper bounds on the maximum grain size that can be reached in extended protostellar envelopes. They considered a coagulation model for the growth of spherical, yet fractal grains whose relative motions are driven by turbulence and the fragmentation of which happens for velocities above a quite generous threshold of v$_{rel}$ = 10 m/s. Considering that the optical properties of dust grains depended on the product of their radius (a) with a filling factor (or porosity) $\phi$, they find that it is not possible to grow grains with an optical radius $a_{opt} = a\phi$ of 1 millimeter, in typical protostellar envelope conditions. They find that, in the $10^5$ yr lifetime of a typical class 0 source whose volume density is approximately $10^7 $ cm$^{-3}$, grains grow up to only about 2.5 $\mu$m.

\subsection{Grain size, composition and temperature effects on $\beta$}
While these theoretical works strongly disfavor the in situ growth of (sub)millimeter dust grains in the envelopes of young objects, the observational evidence of the low dust emissivity indices found therein (\citealt{Kwon09}, \citealt{Miotello2014}, \citealt{Galametz2019}) remains largely uncontested. 

First, an objection to the link between low $\beta$ and large grains might arise from the dependency of the dust-emissivity index on other dust properties.
In fact, the dust emissivity index is sensitive to changes in dust composition, porosity and the presence (or absence) of an icy mantle (e.g., \citealt{BeckwithSargent1991}, \citealt{Testi2014}). Although variations in $\beta$ are evident and well-characterized across different grain properties, $\beta < 1$ remains a strong evidence for the presence of (sub)millimeter grains in the studied distribution ( \citealt{Natta2004ASPC}, \citealt{draine2006}, \citealt{Natta2007}, \citealt{Testi2014} and \citealt{Kholer2015}, \citealt{Ysard2019}).  Thus, although further laboratory studies will be necessary to more precisely interpret $\beta$ in terms of both grain sizes and compositions, dust particles significantly larger than what expected for the typical ISM grains (0.2-2 $\mu$m) remain a robust explanation for our observations.

Secondly, if the compact emission contribution is not properly subtracted from the visibilities, a bias can be introduced when determining the slope of the SED. Previous studies (e.g., \citealt{Miotello2014} and G19) have subtracted the compact emission assuming unresolved disks that would only contribute an offset at all uv-distances. They determined this value by fitting a constant to the visibilities amplitudes after some arbitrary long baseline and subtracted it from the visibilities at shorter uv-distances. Here, we have tried to more properly model and subtract the disk emission as explained in Section \ref{sec:model}.

Finally, perhaps the temperature correction discussed in Section \ref{sec:beta_section} could be revisited. 
If the emission is optically thin, Equation \ref{eq:beta} reduces to:
\begin{equation}
    \beta = \alpha - \frac{d \log(B_{\nu}(T))}{d \log(\nu)},
    \label{eq:corr_beta}
\end{equation}
where the second term is two in case the RJ approximation is satisfied. The temperature power law we consider (cfr. Eq.\ref{eq:temp}) yields temperatures of about 25 K at 200 au, where we find $\beta \lesssim 1$. Here, the correction term of Eq.\ref{eq:corr_beta} amounts to 1.85. If the true temperature of the medium at 300 au were lower than what is predicted with our simple power law, then the real $\beta$ would be higher. However, to produce $\beta \sim$ 1.7, the temperature of the envelope at 300 au would need to be lower than 5 K, which is far below a reasonable value for a protostellar envelope. 

\subsection{Alternatives: differential collapse or outflow transport}
Although simulations predict so far that large grains cannot grow in situ, observations strongly hint to their presence in the inner few hundreds au of these protostellar envelopes. Different mechanisms that can justify their presence might be at play.

First, \citet{Lebreuilly2020} investigated how the distribution of dust grains of a prestellar core evolves during the early phases of the protostellar collapse and how this evolution depends on the initial conditions of the cloud and the dust distribution. They found a significant decoupling between gas and dust for grains of a few 10 $\mu$m. 
Moreover, observations of scattered light from molecular cloud cores in the 3-5 $\mu$m range ("coreshine") have provided evidences for the presence of grains larger than usually expected ISM ones: up to 10 $\mu$m compared to about 0.25 $\mu$m (\citealt{Steinacker2010}, \citealt{Steinacker2014}, \citealt{Steinacker2015}).
These larger grains would tend to settle more efficiently in the first-core, leading to a radial stratification of dust properties that could reproduce a gradient of $\beta$ as we observe here, but hardly explain the low $\beta$ values at the inner envelope scales. Further refinement of this kind of model will require deeper knowledge of prestellar core dust properties.

A second scenario to explain large grains in the envelopes of class 0/I YSO was studied by \citet{Wong2016}, who proposed that such grains might be transported to the envelope after the growth has happened in the disk in the very early stages of the system. Their model suggested that a typical protostellar outflow (T$\sim$ 10 K, v $\sim$ 10 km/s) could lift grains as large as 1 mm in the first $10^4$ yr of the protostellar lifetime if the mass loss rate of the protostar was high enough ($\sim10^{-6} \dot M_{\odot}$/yr). Among others, \citet{Brauer2008}, \citet{Kawasaki2022}, \citet{Bate2022}, \citet{Lebreuilly2023} all find that growth up to cm-sized pebbles is extremely fast (a few $\sim 10^3$ yr) at the high density of the disk environment, thus producing the large grains that the outflow would entrain in the first place.
Based on three-dimensional magneto-hydro-dynamical simulations, \citet{Lebreuilly2020} and \citet{Tsukamoto2021} found that the large dust grains grown in the inner region of a disk can be entrained by an outflow up to the envelope scales (up to 100 $\mu$m according to the first work, and even $>$1 mm for the second). These grains then decouple from the gas and are ejected from the outflow into the envelope itself, enriching its dust population before falling back to the disk. 

\citet{chen16} found evidence for low $\beta$ correlated with protostellar outflow locations in Perseus. More recently, G19 observed a correlation between $\beta$ and the envelope mass of the CALYPSO sources, which might be in turn caused by a correlation between envelope mass and CO outflows fluxes \citep{Bontemps1996}.  
This two relationships together might support the scenario of disk dust transport into class 0/I protostellar envelopes.

While a more thorough treatment of this link for L1527 is outside of the scope of this work, we note that L1527 indeed hosts a large scale outflow structure in the east-west direction, perpendicular to the edge-on disk. \citet{Hogerheijde1998} observed the source with the James Clerk Maxwell Telescope and detected in $^{12}$CO ($J$=3–2) line emission an outflow extending over about 20000 au. Moreover, \citet{Tobin2010a} imaged this source with the Gemini North telescope (in its L' band) and with the Infrared Array Camera (IRAC) on the Spitzer Space Telescope (2.15-8.0 $\mu$m), detecting the outflow cavity out to roughly 10000 au in both cases. Future investigations should consider whether the L1527 outflow can transport (sub)millimeter grains from the inner disk and lift them in the envelope. Such research is vital to our understanding the origin and consequences of relatively large grains at envelope scales, and may even shed light on this evolutionary stage of disk-based planet formation models.

\subsection{Caveats}
\label{section:caveats}

Although we aim to present a more robust way of measuring the spectral index of the faint extended emission from the envelope of class 0/I YSOs, our analysis is not free of caveats. Additional work remains to further improve the reliability of our claims.

First, while we present a study of the spectral- and dust emissivity indices as a function of uv-distance and physical source scales, we have to keep in mind that a particular baseline does not only probe an annular region at some distance from the center, rather it probes any physical structure within a scale given by $\theta \sim \lambda/B$. It is for this reason that we have to model and subtract the disk from the visibility amplitudes before evaluating the spectral index for the envelope and, for the same reason, the $\alpha$ and $\beta$ we compute for the envelope at different scales represent a flux-weighted average over the relevant envelope spatial scales. If the inner envelope is much brighter than the outer one, as expected, the flux at even the shorter baselines will be dominated by the inner envelope. Thus, the spectral index value at these scales would still be mainly describing the inner region. 

Second, in Section \ref{section:spindx} we have calculated the dust emissivity index $\beta$ of the envelope taking into account diskrepancies from the RJ approximation, hence accounting for a temperature profile, T $\propto$ R$^{-0.4}$. This power-law was derived for a dusty envelope illuminated by a central protostar (e.g., \citealt{Terebey1993}). While the values we obtain using this proportionality are reasonable, as they do not reach values lower than typical cores temperatures at typical cores scales (10 K at 2000 au; e.g., \citealt{Ferriere2001}), a more thorough treatment that accounts for the temperature structure and its associated uncertainties can be achieved by post-processing 3D envelope models with radiative transfer in order to. 

\section{Conclusions}
We aim to characterize the maximum grain size of the dust distribution in the envelope of a class 0/I YSO: L1527 IRS, or L1527 for short. Given its vicinity (140 pc), this source has been extensively studied over decades with different types of data and methodologies to investigate its star, extended envelope, and circumstellar disk both in the dust continuum and line emission. In this context, we exploit the richness of data from the Atacama Large (sub)millimeter Array to greatly enhance the S/N of the extended emission in the continuum and determine the peak of the dust grain population in the envelope that surrounds and infalls onto the young circumstellar disk of L1527. 

We find that:
\begin{itemize}
    \item[$\bullet$] The spectral index of the compact disk, i.e. within a radius of $\sim$ 75 au, is $\alpha_{0.88-3.4mm} = 2$, consistent with the expected high optical depth of its edge-on geometry. No dust grain properties information can be extracted in such a condition without thorough radiative transfer modelling, outside of the scope of this work.
    \item[$\bullet$] After subtraction of the inner disk emission and correction for the RJ approximation, we find tentative evidence for a positive outward gradient of the dust emissivity index of the envelope of L1527 ($\beta(R) \sim 0.37 \log$(R$_{au}$), which can be interpreted as variations of dust properties from ISM-like particles ($\beta \sim$ 1.6) at 2000 au down to grown grains below 300 au ($\beta \leq $1).
    \item[$\bullet$]  The implications of physical and chemical properties of dust grains on the values of $\beta$ as well as the possible impact of the temperature profile used to calculate the correction for departure from the RJ approximation in computing the dust emissivity index (cfr Section \ref{section:beta_discuss}) is carefully discussed. None of these possibilities alone are able to justify $\beta$ values as low as 1 at hundreds of au, thus strengthening the hypothesis of submillimeter grains at these scales. While in situ formation seems disfavored by most theoretical studies, differential core dust collapse or outflow dust transport from the disk could explain our observations.
    \item[$\bullet$] We argue that a multiscale ($10^1-10^3$ au), multifrequency (n>2), and high-sensitivity study is necessary to tightly constrain the spectral- and dust emissivity index profiles of faint class 0/I YSO envelopes. Using ALMA B3, B4, B6 and B7 effectively halves the error bars of measurements made in previous studies that made use of two wavelengths only.
\end{itemize}

ALMA has provided us with the necessary capabilities in terms of resolution, recoverable scales, sensitivity, and frequency ranges that are critical to the study of the continuum emission of faint, extended envelopes such as the one infalling onto the young protostar L1527. It is now the time to exploit this infrastructure to conduct sample studies of these objects with state-of-the-art data sets to finally pinpoint the initial conditions of grain growth and planet formation in both space and time.

\begin{acknowledgements}
This work was partly supported by the Italian Ministero dell’Istruzione, Universit\`{a} e Ricerca through the grant Progetti Premiali 2012-iALMA (CUP C52I13000140001), by the Deutsche Forschungsgemeinschaft (DFG, German Research Foundation) - Ref no. 325594231 FOR 2634/2 TE 1024/2-1, by the DFG Cluster of Excellence Origins (www.origins-cluster.de). This project has received funding from the European Union’s Horizon 2020 research and innovation program under the Marie Sklodowska- Curie grant agreement No 823823 (DUSTBUSTERS) and from the European Research Council (ERC) via the ERC Synergy Grant ECOGAL (grant 855130). This research has received funding from the European Research Council (ERC) under the European Union’s Horizon 2020 research and innovation programme (MagneticYSOS project, Grant Agreement No 679937). D.J.\ is supported by NRC Canada and by an NSERC discovery Grant. We thank the entire ALMA team for their dedication to provide us with the data we used for this work. This paper makes use of the following ALMA data: ADS/JAO.ALMA\#2015.1.00261.S, ADS/JAO.ALMA\#2016.1.01245.S, ADS/JAO.ALMA\#2016.1.01541.S, ADS/JAO.ALMA\#2017.1.00509.S, ADS/JAO.ALMA\#2018.1.01205.L, ADS/JAO.ALMA\#2016.1.01203.S, ADS/JAO.ALMA\#2012.1.00647.S, ADS/JAO.ALMA\#2013.1.01086.S, ADS/JAO.ALMA\#2012.1.00193.S, ADS/JAO.ALMA\#2017.1.01413.S, ADS/JAO.ALMA\#2011.0.00604.S, ADS/JAO.ALMA\#2011.0.00210.S, ADS/JAO.ALMA\#2013.1.01331.S, ADS/JAO.ALMA\#2012.1.00346.S, ADS/JAO.ALMA\#2015.1.01549.S, ADS/JAO.ALMA\#2016.A.00011.S, ADS/JAO.ALMA\#2018.1.00799.S, ADS/JAO.ALMA\#2016.2.00171.S, ADS/JAO.ALMA\#2016.2.00117.S. ALMA is a partnership of ESO (representing its member states), NSF (USA) and NINS (Japan), together with NRC (Canada), MOST and ASIAA (Taiwan), and KASI (Republic of Korea), in cooperation with the Republic of Chile. The Joint ALMA Observatory is operated by ESO, AUI/NRAO and NAOJ. We thank M. Troisi for his sharp comments, which supported LC during this work. We thank the referee for their comments, which helped to improve the quality of this work. 
\end{acknowledgements}

\bibliographystyle{aa} 
\bibliography{biblio} 

\begin{appendix}
\section{Continuum maps}
While we have not conducted our analysis on the continuum maps, we have produced a set of images for each band during self-calibration, separately for the compact and extended array (see Section \ref{sec:data_red}). We here report both large-scale (ACA 7m-array) and higher resolution (ALMA 12m-array) continuum maps of L1527 for the bands we used in this work.
\begin{figure}[h]
    \centering
    \includegraphics[width=\linewidth]{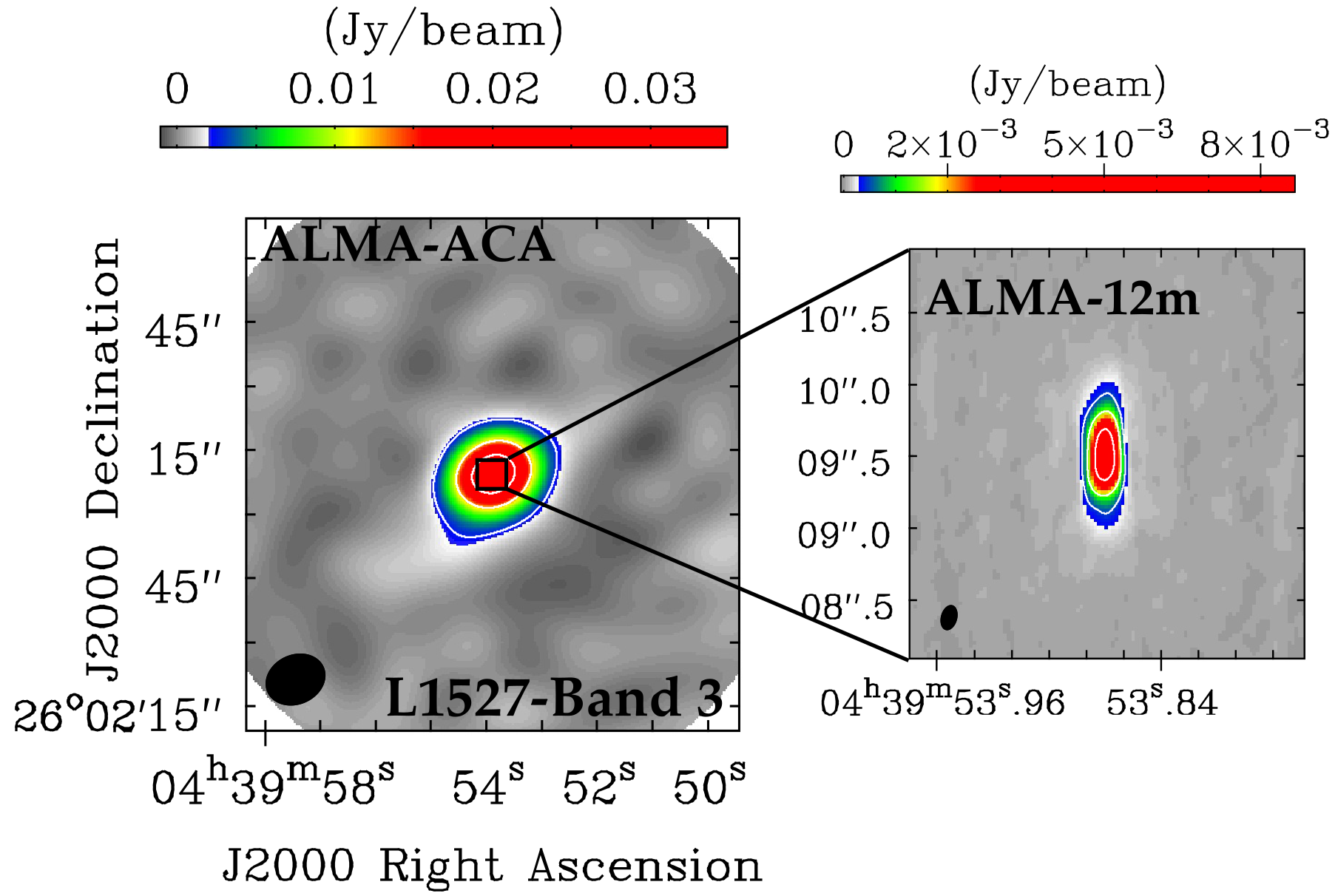}
    \caption{ALMA B3 continuum map of L1527 obtained imaging the 7m-array setups only (left) and the 12m-array only (right). 
    The color map for the ACA image shows only flux densities higher than 10$\sigma$, and the white [50, 150]$\sigma$ level contours.
    The color map of the 12m-array image shows only flux densities higher than 50$\sigma$, and the white [150, 500]$\sigma$ level contours. 
    The synthesized beams are shown in black in the lower left corner in both cases. We note that the colorbar is different for each map.}
    \label{fig:ALMA_B3}
\end{figure}
\begin{figure}[h]
    \centering
    \includegraphics[width=\linewidth]{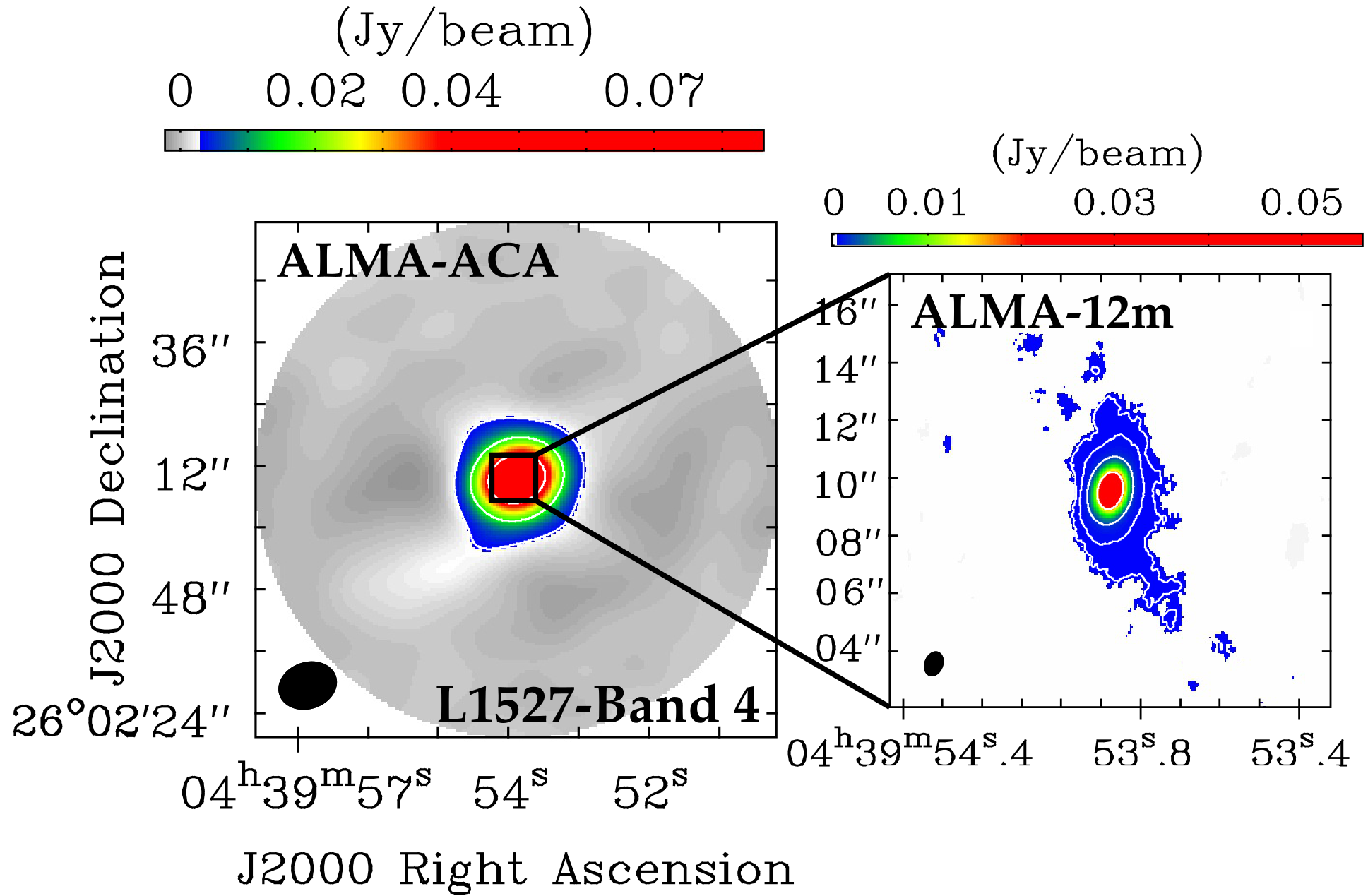}
    \caption{ALMA B4 continuum map of L1527 obtained imaging the 7m-array setups only (left) and the 12m-array only (right). 
    The color map for the ACA image shows only flux densities higher than 10$\sigma$, the the white [50, 150]$\sigma$ level contours.
    The color map of the 12m-array image shows only flux densities higher than 5$\sigma$, and the white [5, 10, 50, 500]$\sigma$ level contours. 
    The synthesized beams are shown in black in the lower left corner in both cases. We note that the colorbar is different for each map.}
    \label{fig:ALMA_B4}
\end{figure}
\begin{figure}[h]
    \centering
    \includegraphics[width=\linewidth]{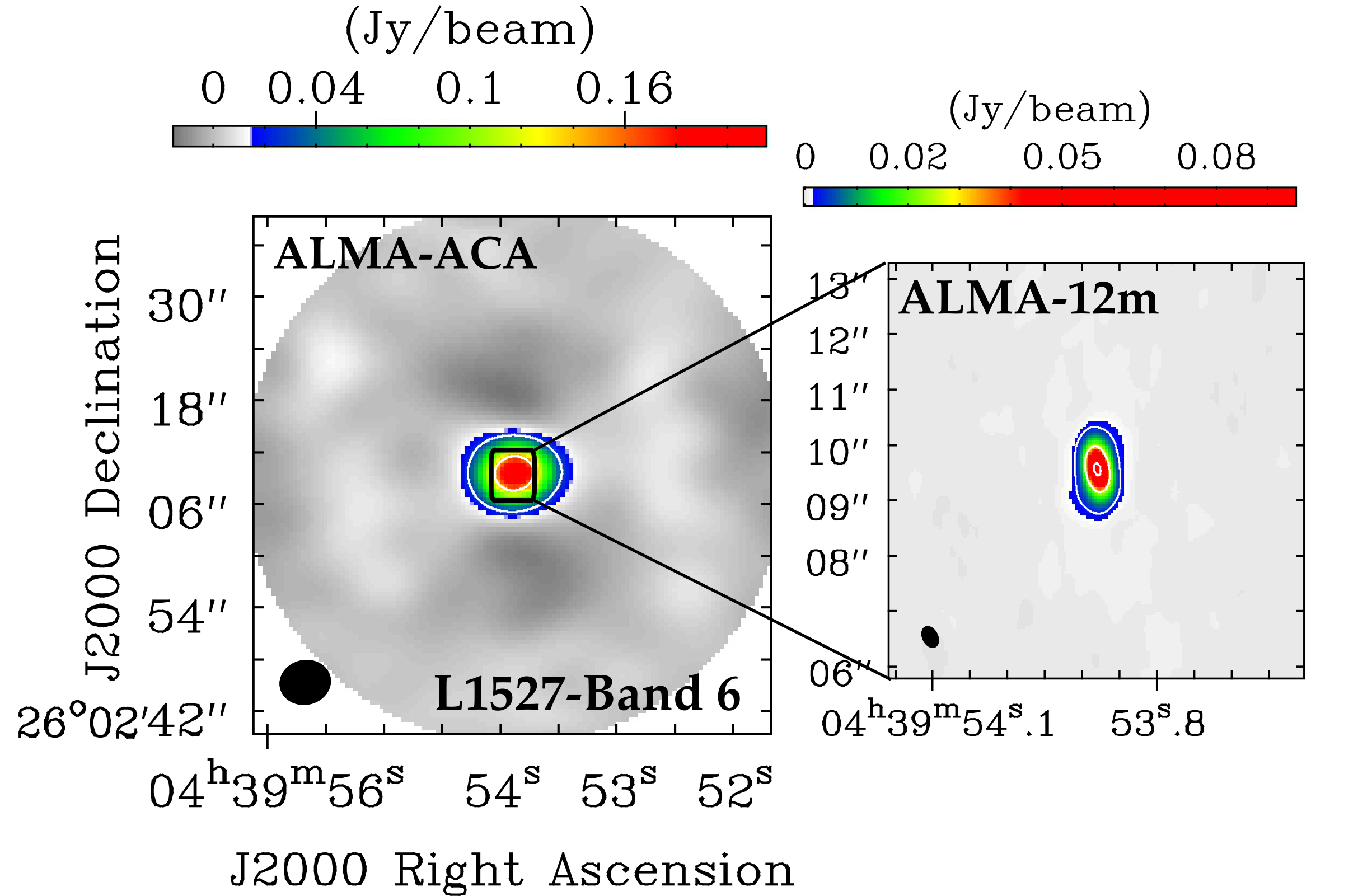}
    \caption{ALMA B6 continuum map of L1527 obtained imaging the 7m-array setups only (left) and the 12m-array only (right). 
    The color map for the ACA image shows only flux densities higher than 10$\sigma$, and the white [50, 150]$\sigma$ level contours.
    The color map of the 12m-array image shows only flux densities higher than 5$\sigma$, and the white [5, 50, 150]$\sigma$ level contours. 
    The synthesized beams are shown in black in the lower left corner in both cases. We note that the colorbar is different for each map.}
    \label{fig:ALMA_B6}
\end{figure}
\begin{figure}[h]
    \centering
    \includegraphics[width=\linewidth]{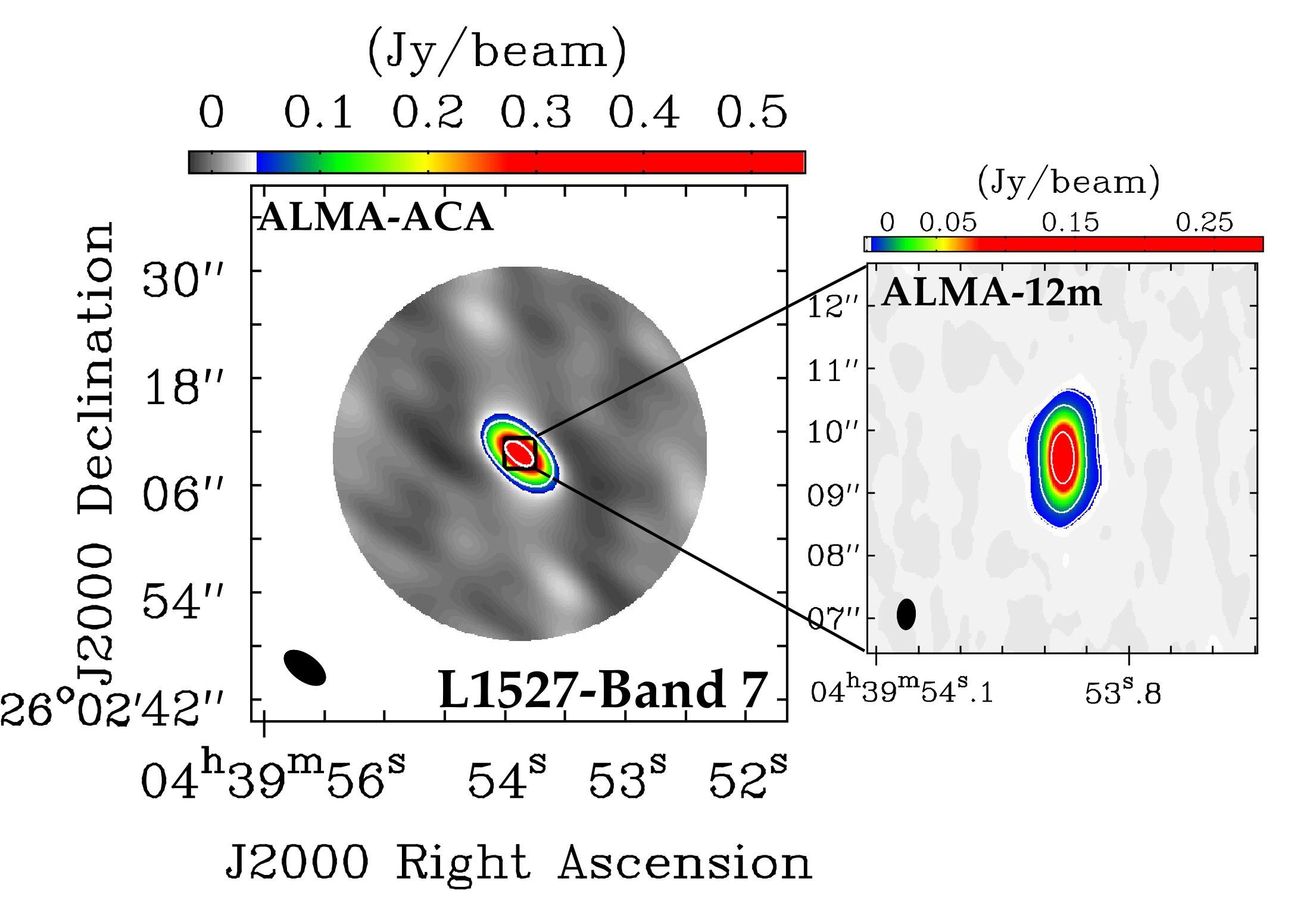}
    \caption{ALMA B7 continuum map of L1527 obtained imaging the 7m-array setups only (left) and the 12m-array only (right). 
    The color map for the ACA image shows only flux densities higher than 10$\sigma$, and the white [10, 50]$\sigma$ level contours.
    The color map of the 12m-array image shows only flux densities higher than 10$\sigma$, and the white [10, 50, 500]$\sigma$ level contours. 
    The synthesized beams are shown in black in the lower left corner in both cases. We note that the colorbar is different for each map.}
    \label{fig:ALMA_B7}
\end{figure}

\section{Adjacent bands spectral index}
\label{sec:appendix1}
In this study, we have used multiple wavelengths to study the continuum emission of the envelope of L1527. This is critical to damp the possible systematic errors that occur when only using fluxes at two wavelengths. We show the importance of this approach in what follows.
We computed $\alpha$ between pair of bands as follows:
\begin{equation}
    \alpha = \frac{log_{10}{F(\nu_2)}-log_{10}{F(\nu_1)}}{log_{10}{\nu_2}-log_{10}{\nu_1}}.
    \label{eq:alpha}
\end{equation}
To evaluate the uncertainties on the spectral index between two wavelengths, we propagate the errors of Eq.\ref{eq:alpha}:
\begin{equation}
    \Delta\alpha^2 = \Bigg(\frac{1}{ln{\nu_2}-ln{\nu_1}}\Bigg)^2 \Bigg(\frac{\sigma_1^2}{F_{\nu_1}^2} + \frac{\sigma_2^2}{F_{\nu_2}^2}  \Bigg),
\end{equation}

\begin{figure}[ht]
    \centering
    \includegraphics[width=\linewidth]{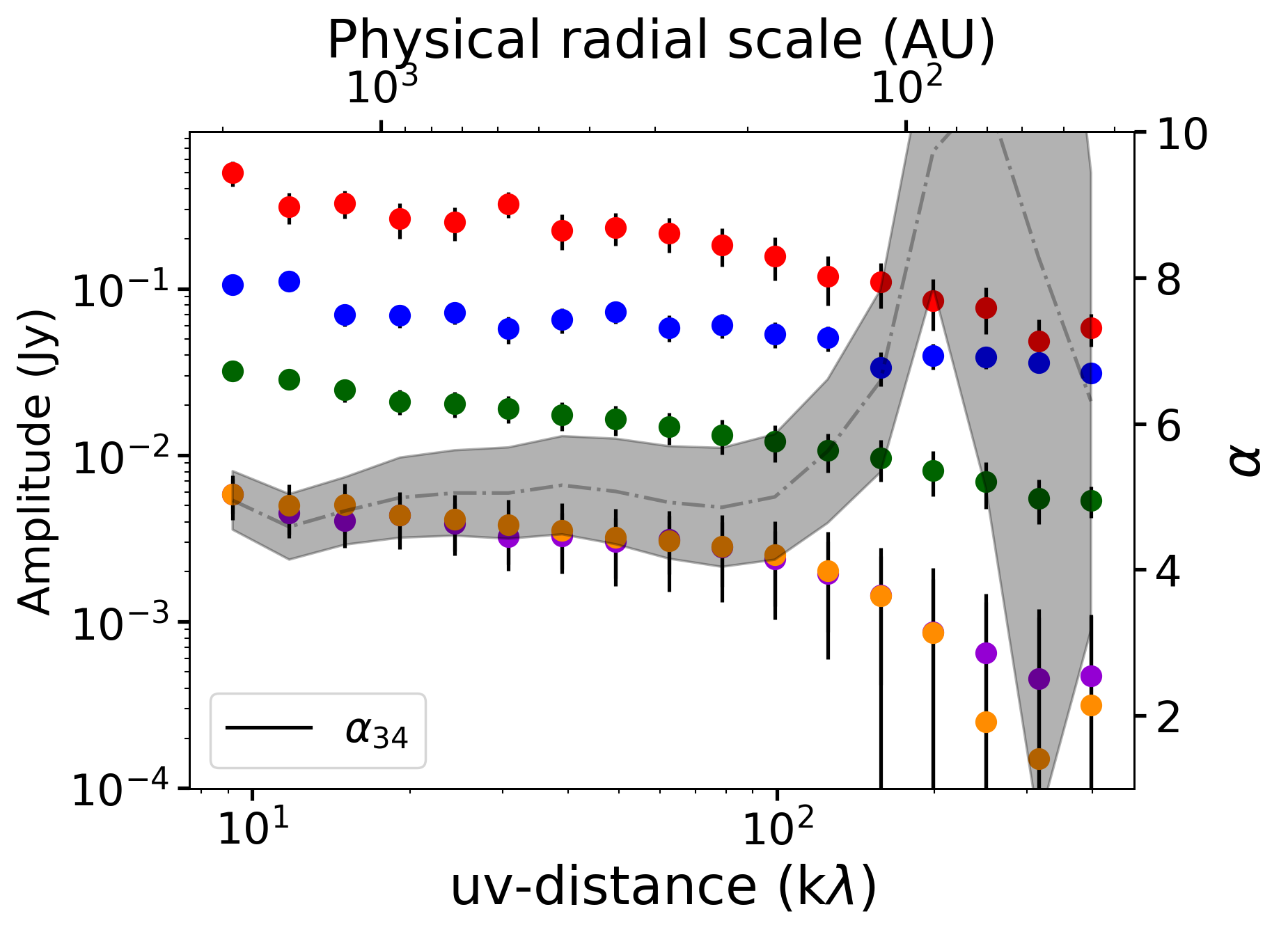}
    \caption{Same as Fig.\ref{fig:mean_alpha} but for B3 (100 GHz) - B4 spectral index.}
    \label{fig:B34}
\end{figure}
\begin{figure}[ht]
    \centering
    \includegraphics[width=\linewidth]{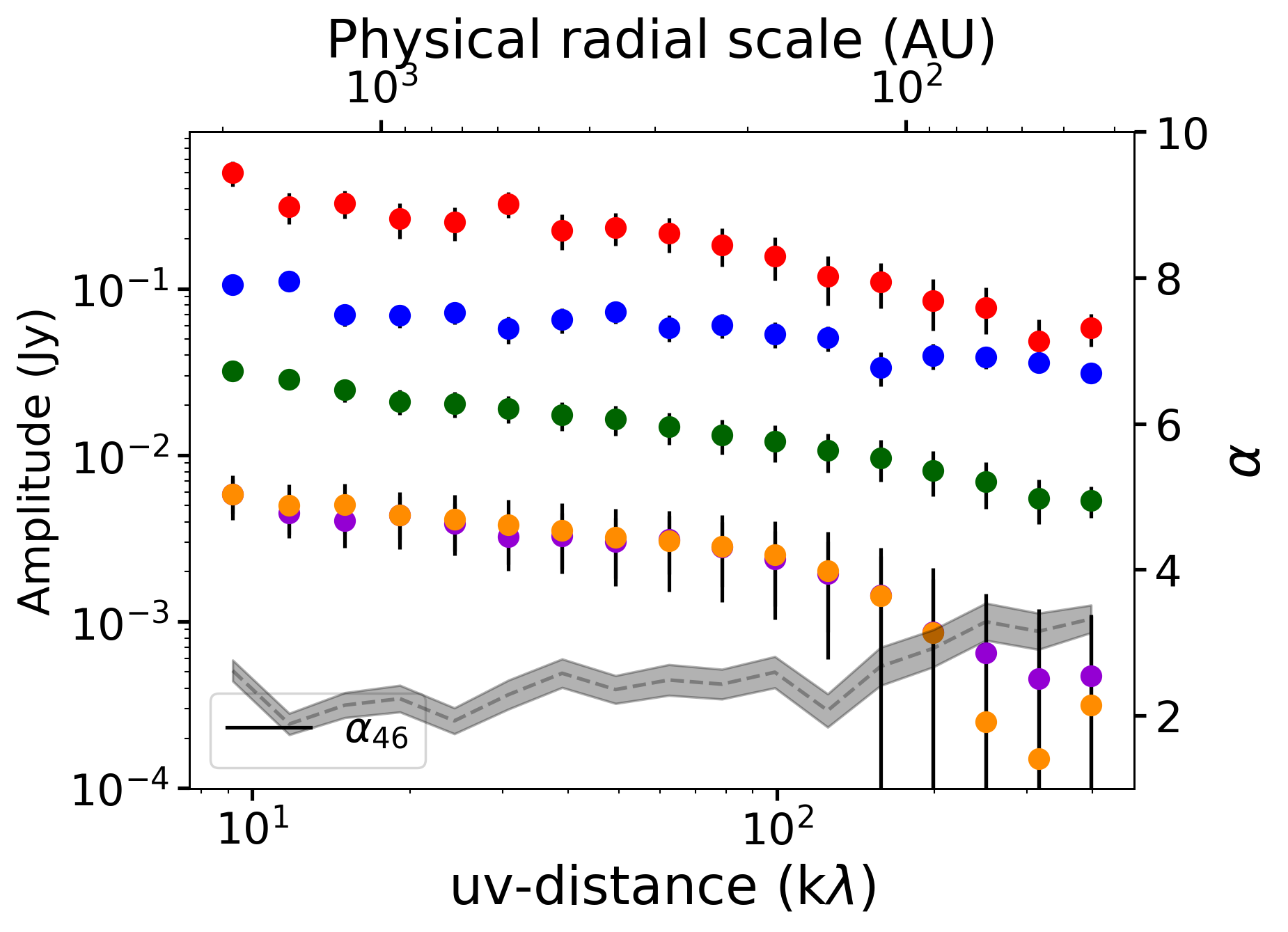}
    \caption{Same as Fig. \ref{fig:B34} but for B4 and B6.}
    \label{fig:B46}
\end{figure}
\begin{figure}[ht]
    \centering
    \includegraphics[width=\linewidth]{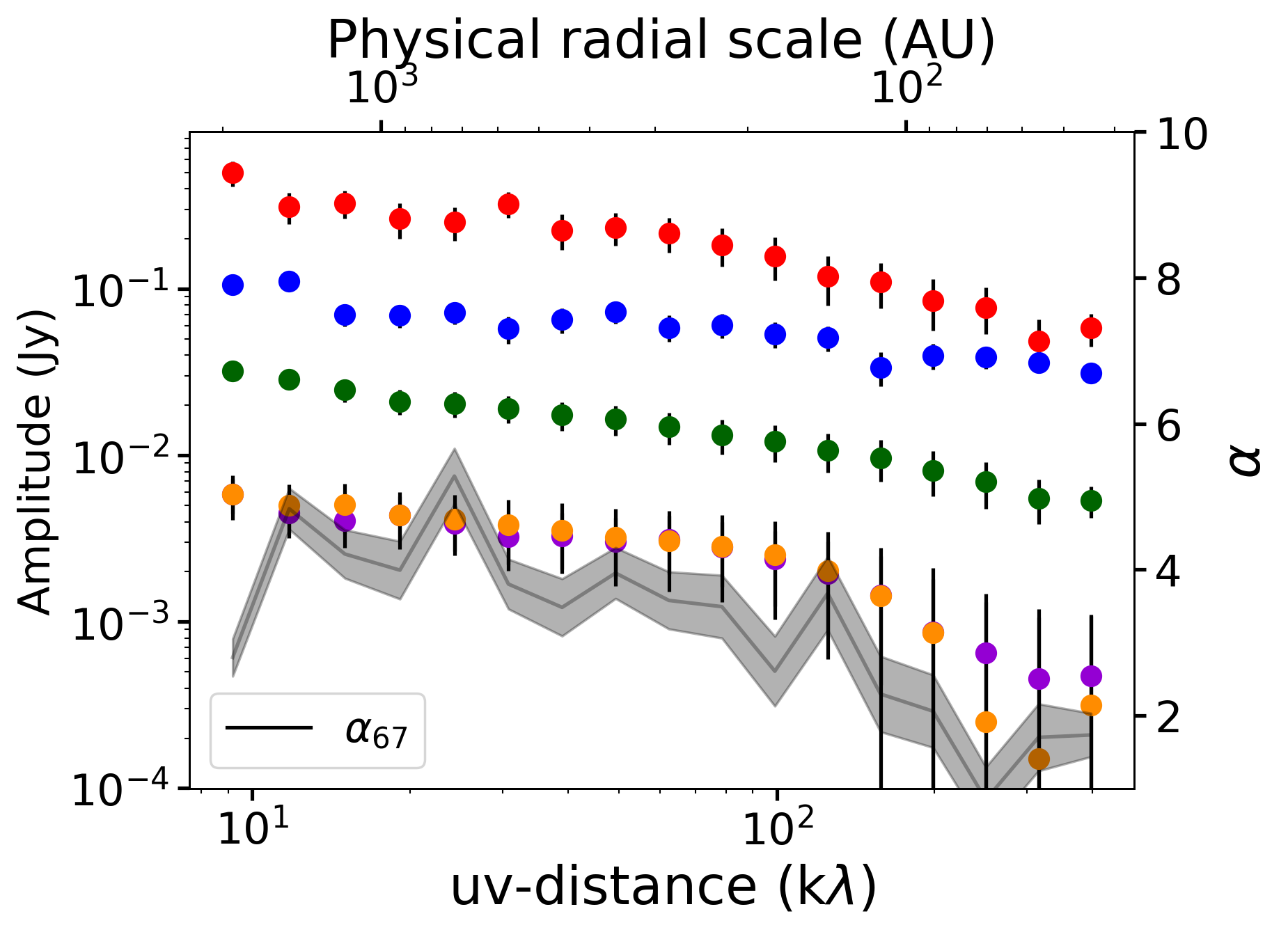}
    \caption{Same as Fig. \ref{fig:B34} but for B6 and B7.}
    \label{fig:B67}
\end{figure}
\begin{figure}[ht]
    \centering
    \includegraphics[width=\linewidth]{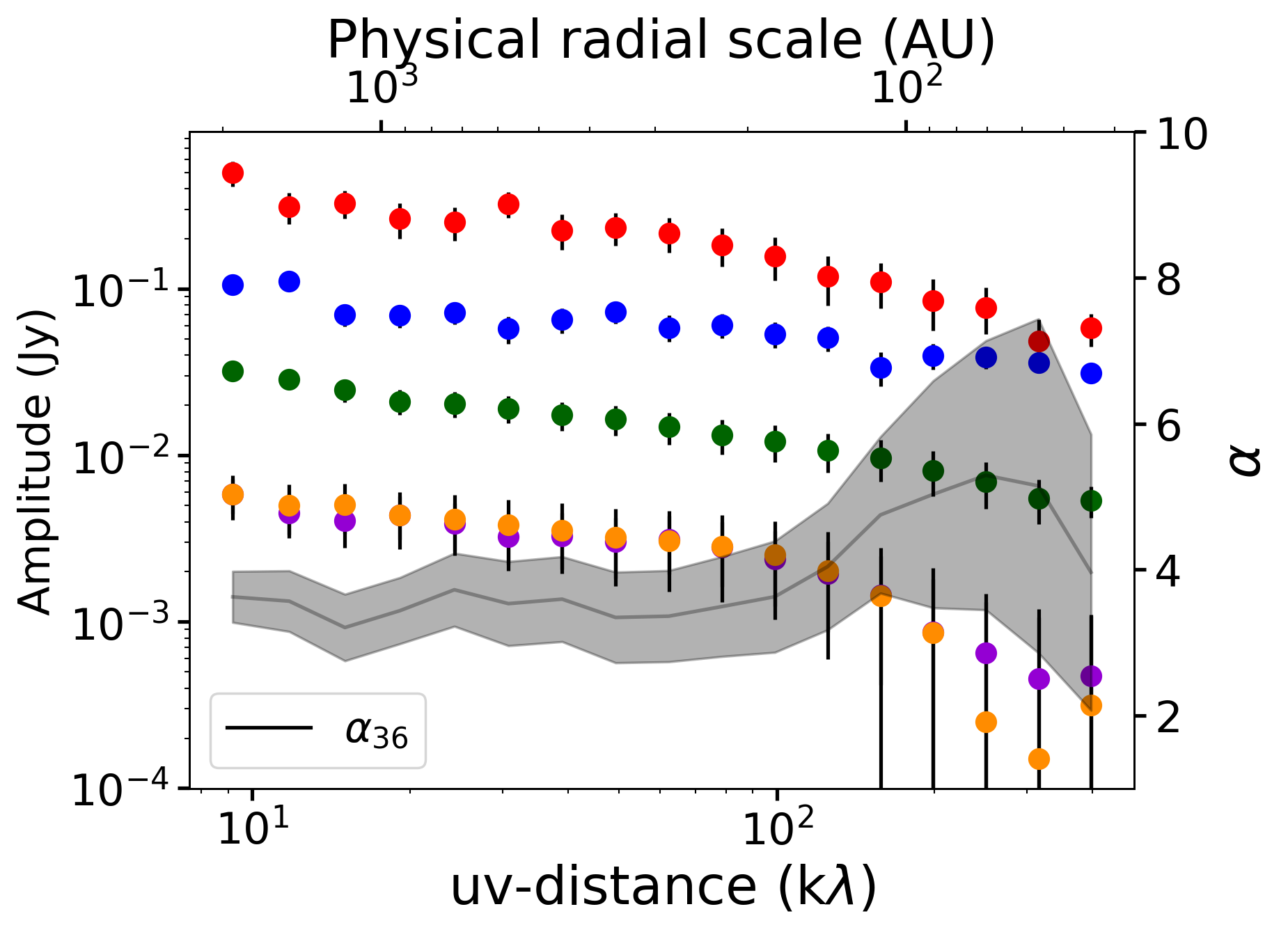}
    \caption{Same as Fig. \ref{fig:B34} but for B3 and B6.}
    \label{fig:B36}
\end{figure}

where $\sigma_1$ and $\sigma_2$ are the uncertainties on the amplitudes in each uv-distance bin:
\begin{equation}
    \sigma_i = \sqrt{ \Bigg(\frac{\partial F_i}{\partial Re_i} \Bigg)^2  dRe_i^2 +\Bigg(\frac{\partial F_i}{\partial Im_i} \Bigg)^2 dIm_i^2 + \Bigg(F_i \cdot dC \Bigg)^2}.
    \label{eq:amp_err}
\end{equation}
Here, ${Re}$ and ${Im}$ are the real and imaginary parts of the interferometric visibilities that make the amplitude as $F = \sqrt(Re^2 + Im^2)$, and $dC$ is the flux calibration error. We set the latter to 10\% for B7 and 5\% for  B3, B4 and B6, following the prescriptions of the ALMA Handbook (\citealt{Remijan2019}, \citealt{francis20}). 
We here show the spectral index of the envelope of L1527 as computed between adjacent bands, plus the one between B3 and B6. Figures \ref{fig:B34}, \ref{fig:B46}, \ref{fig:B67}, \ref{fig:B36} show how, given the statistical errors and the proximity of the adjacent bands, the error on $\alpha$ can be large at the longest baselines. 
Moreover, the spectral index $\alpha$ shows systematic differences among different combination of adjacent bands. Only using multiple bands (>2), one can robustly constrain the spectral index, damping the uncertainties that arise using only two frequencies (see Section \ref{section:spindx}).

\section{Fit results}
\label{sec:appendix2}
\begin{figure}[h]
    \centering
    \includegraphics[width=0.75\linewidth]{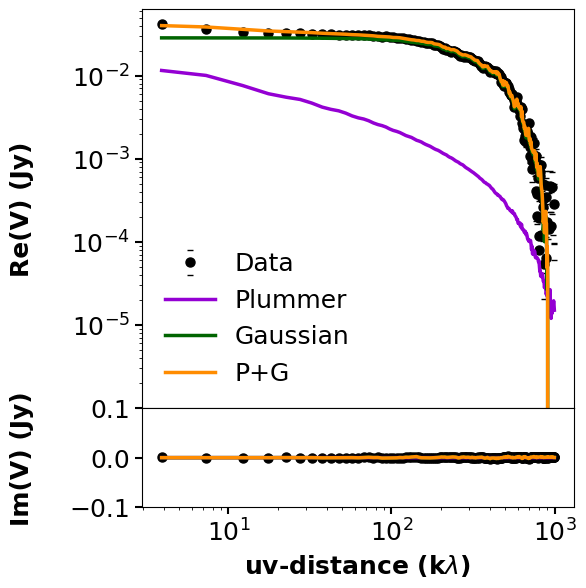}
    \caption{The \texttt{galario} fit of the B3 (100 GHz) real and imaginary part of the visibilities. The best model (orange) is composed of a compact Gaussian model (green) and an outer power law (violet). The wiggles in the plotted model are due to its sampling on the uv points of the observations.}
    \label{fig:B3fit}
\end{figure}

\begin{figure}
    \centering
    \includegraphics[width=0.75\linewidth]{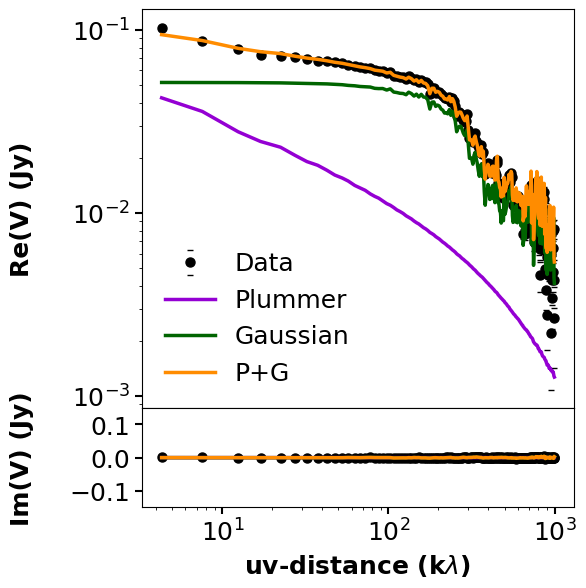}
    \caption{Same as Fig. \ref{fig:B3fit} but for B4.}
    \label{fig:B4fit}
\end{figure}

\begin{figure}
    \centering
    \includegraphics[width=0.75\linewidth]{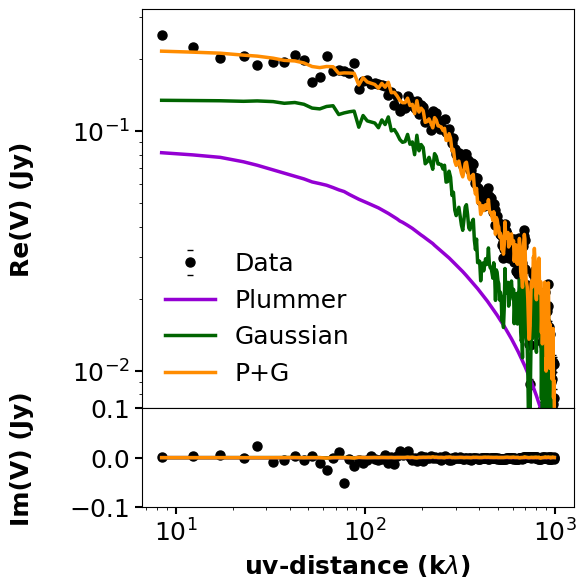}
    \caption{Same as Fig. \ref{fig:B3fit} but for B6.}
    \label{fig:B6fit}
\end{figure}

\begin{figure}
    \centering
    \includegraphics[width=0.75\linewidth]{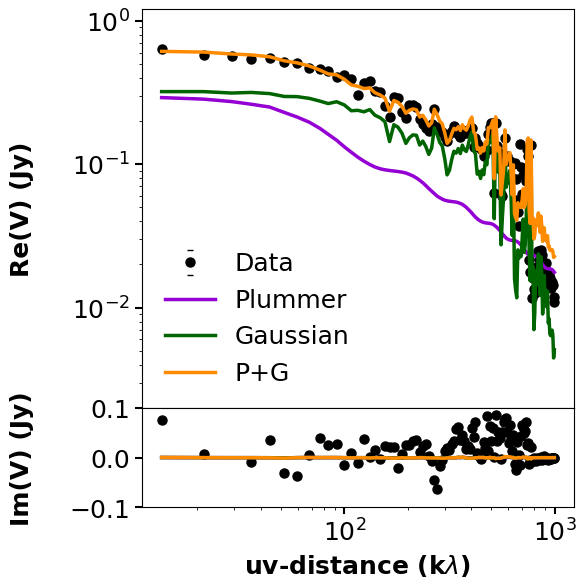}
    \caption{Same as Fig. \ref{fig:B34} but for B7.}
    \label{fig:B7fit}
\end{figure}

As explained in Section \ref{sec:model}, we fit a composite model to the visibilities. The model consists of a Gaussian component to trace the compact emission and a power law to better describe the extended emission. The disk radius, defined as the $2\sigma$ level of the Gaussian component is approximately 75 au, consistent with what was found kinematically by \citet{Aso2017}. The disk contributes to the flux up to a maximum of 98\% in B3, highlighting the necessity to combine as many ALMA datasets as possible and selfcalibrate them in order to enhance the faint extended continuum emission. 
The disk results edge-on ($inc \sim 80^{\circ}$), consistently with several results throughout the literature with a position angle $PA \sim 2^{\circ}$, in a westward convention. The inner radius of the envelope is constrained to be in the range 0.01-0.15 arcseconds while the outer one is in the range 8-12 arcseconds, both depending on the band. Having phase-centered the datasets during the data calibration procedure, we find both dRA and dDec phase offsets consistent with zero. Finally, the power law exponent of the Plummer profile ($p$) is found to be in the range 2.6-3.1, roughly consistent with what was found by \citet{Ladd1991}. We list the results of the fitting, that was carried out with \textsf{galario}, in Tab.\ref{tab:fits_results} and show the final model in Figures \ref{fig:fitband3}, \ref{fig:B3fit}, \ref{fig:B4fit}, \ref{fig:B6fit}, \ref{fig:B7fit}.

\begin{table*}[h!]
\begin{center}
\renewcommand{\arraystretch}{2}
\begin{tabular}{|c|ccccc|}
\hline
Parameter & Band 3  (88 GHz)    & Band 3 (100 GHz) & Band 4 & Band 6 & Band 7 \\  \hline 
          
$f_0$ (Jy/sr)    & $10.47 \pm 0.02$  & $10.53 \pm 0.01$  & $10.79 \pm 0.03$     & $11.12 \pm 0.01$ &  $11.51 \pm 0.01$      \\ \hline 
$r$       & $0.90 \pm 0.01$   &$0.98 \pm 0.01$    & $0.74 \pm 0.01$      &$0.72 \pm 0.01$ &  $0.38 \pm 0.01$      \\ \hline
$\sigma$ (") & $0.147 \pm 0.007$ &$0.151 \pm 0.001$  &$0.208 \pm 0.004$     &$0.270 \pm 0.001$ &   $0.330 \pm 0.001$     \\ \hline
inc  (deg)     & $78 \pm 1$       &$80 \pm 1$      &$80 \pm 1$      &$82\pm1$ &     $78 \pm 1$   \\ \hline
PA   (deg)     & $2.08\pm0.6$        &$2.27 \pm 0.02$      & $3.09 \pm 0.05$        &$2.3\pm0.01$ &     $1.942 \pm 0.007$   \\ \hline
$R_{in}$ (")  & $0.03\pm0.01$   &$0.14 \pm 0.01$     & $0.04 \pm 0.01$      &$0.08 \pm 0.02$ &   $0.03\pm 0.01$     \\\hline
$R_{out}$ (") & $9.9\pm0.1$       &$9.9 \pm 0.1$      & $12.8 \pm 0.1$         &$4.4 \pm 0.2$ &     $1.35 \pm 0.1$   \\ \hline 
$p$       & $2.62\pm0.01$     &$2.76 \pm 0.03$      & $2.75 \pm 0.01$      & $3.37 \pm 0.01$ &   $2.54 \pm 0.02$     \\ \hline 
dRA (")     & $0.029\pm0.001$ &$0.029 \pm 0.001$  & $0.007 \pm 0.001$  &$0.001 \pm 0.001$ &  $0.005 \pm 0.001$      \\ \hline 
dDec (")    & $-0.017\pm0.002$  &$-0.018 \pm 0.001$ & $-0.011 \pm 0.001$ & $0.008 \pm 0.005$ & $0.005 \pm 0.003$     \\ \hline 

\end{tabular}
\end{center}
\label{tab:fits_results}
\caption{Best-fit parameters of the Gaussian plus power law model as obtained with \textsf{galario}, for each ALMA Band.}
\end{table*}

\end{appendix}

\end{document}